\documentclass[lettersize,journal]{IEEEtran}
\usepackage{amsmath,amsfonts}
\usepackage{algorithmic}
\usepackage{algorithm}
\usepackage{array}
\usepackage[caption=false,font=normalsize,labelfont=sf,textfont=sf]{subfig}
\usepackage{textcomp}
\usepackage{stfloats}
\usepackage{url}
\usepackage{verbatim}
\usepackage{graphicx}
\usepackage{cite}
\usepackage{multirow}
\usepackage{makecell}
\usepackage{threeparttable}
\begin{document}

\title{SpeechFormer++: A Hierarchical Efficient Framework for Paralinguistic Speech Processing}
\author{Weidong Chen, Xiaofen Xing, Xiangmin Xu, Jianxin Pang, and Lan Du
\thanks{This work was supported in part by the National Key R\&D Program of China under Grant 2022YFB4500600; in part by the National Natural Science Foundation of China under Grant U1801262; in part by the Science and Technology Project of Guangzhou under Grant 202103010002; in part by the Science and Technology Project of Guangdong under Grant 2022B0101010003; in part by the Natural Science Foundation of Guangdong Province under Grant 2022A1515011588; and in part by the Guangdong Provincial Key Laboratory of Human Digital Twin under Grant 2022B1212010004. \textit{(Corresponding authors: Xiangmin Xu; Xiaofen Xing.)}}
\thanks{Weidong Chen and Xiaofen Xing are with the School of Electronic and Information Engineering, South China University of Technology, Guangzhou 510640, China (e-mail: eewdchen@mail.scut.edu.cn; xfxing@scut.edu.cn). Xiangmin Xu is with the School of Future Technology, South China University of Technology, Guangzhou 511442, China, and also with Pazhou Laboratory, Guangzhou 510330, China (e-mail: xmxu@scut.edu.cn). Jianxin Pang is with UBTECH Research, UBTECH Robotics Corporation, Shenzhen 518055, China (e-mail: walton@ubtrobot.com). Lan Du is with iFLYTEK Research, iFLYTEK Corporation, Hefei 230088, China (e-mail: landu@iflytek.com).}
}

\markboth{IEEE/ACM TRANSACTIONS ON AUDIO, SPEECH, AND LANGUAGE PROCESSING}
{Shell \MakeLowercase{\textit{et al.}}: A Sample Article Using IEEEtran.cls for IEEE Journals}

\maketitle

\begin{abstract}
Paralinguistic speech processing is important in addressing many issues, such as sentiment and neurocognitive disorder analyses. Recently, Transformer has achieved remarkable success in the natural language processing field and has demonstrated its adaptation to speech. However, previous works on Transformer in the speech field have not incorporated the properties of speech, leaving the full potential of Transformer unexplored. In this paper, we consider the characteristics of speech and propose a general structure-based framework, called SpeechFormer++, for paralinguistic speech processing. More concretely, following the component relationship in the speech signal, we design a unit encoder to model the intra- and inter-unit information (\textit{i.e.}, frames, phones, and words) efficiently. According to the hierarchical relationship, we utilize merging blocks to generate features at different granularities, which is consistent with the structural pattern in the speech signal. Moreover, a word encoder is introduced to integrate word-grained features into each unit encoder, which effectively balances fine-grained and coarse-grained information. SpeechFormer++ is evaluated on the speech emotion recognition (IEMOCAP \& MELD), depression classification (DAIC-WOZ) and Alzheimer's disease detection (Pitt) tasks. The results show that SpeechFormer++ outperforms the standard Transformer while greatly reducing the computational cost. Furthermore, it delivers superior results compared to the state-of-the-art approaches.

\end{abstract}


\section{Introduction}

\IEEEPARstart{S}{peech} has been used over thousands of years in human society and is able to convey the most information in the simplest way\cite{perception}. Paralinguistic speech processing (PSP), which aims to extract information beyond the linguistic content of speech, such as sentiment, depression and neurocognition, has a wide range of applications in different areas. Consequently, it is of increasing interest to the research community.

Modeling speech signals in PSP is a great challenge because the pronunciation information and the dynamic changes of speech are well understood by humans but are difficult  for models to comprehend. Over the last three decades, numerous machine learning algorithms, such as hidden Markov models \cite{HMM1,HMM2,HMM3}, decision trees \cite{Tree2,Tree3} and restricted Boltzmann machines \cite{Boltzmann2,Haizhou3,Boltzmann3}, have been proposed to capture paralinguistic information in speech. Recently, deep learning methods have delivered superior performance for PSP tasks owing to their remarkable modeling capabilities. For example, convolutional neural networks (CNNs) \cite{GCNN, Makiuchi, EmoAudioNet, FVTC-CNN, stc, ISNet, fan2020}, graph neural networks (GNNs) \cite{LSTM-GIN, MM-DFN}, recurrent neural networks (RNNs) \cite{MMFA-RNN, Autoencoder, SER_RNN1} and two popular variants of the RNNs named long short-term memory (LSTM) \cite{Tao1, tao3, Romain1} and gated recurrent units (GRUs) \cite{SER_RNN3} have achieved promising results in PSP domain.

\begin{figure}[t]
\centering
\includegraphics[width=3.4in]{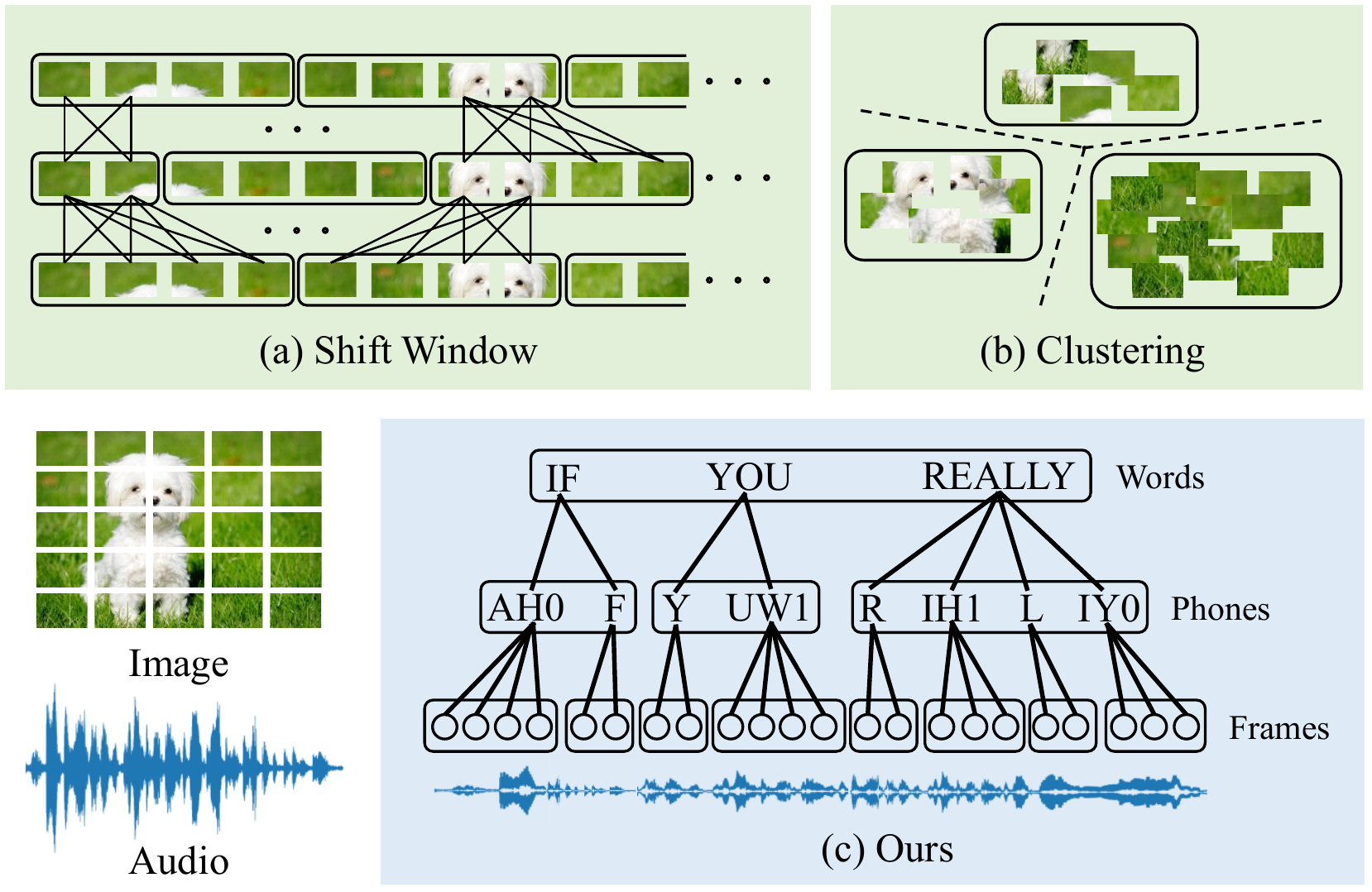}
\caption{(a) (b) Graphical illustration of two typical attention variants in computer vision domain and (c) the proposed attention mechanism for speech signal. Each rectangle represents a local attention window.
}
\label{fig_1}
\end{figure}

Transformer \cite{Transformer}, is the latest deep learning technique and was originally used in the natural language processing field (NLP). It has achieved great success using full attention to model the long-range dependencies in a sequence. Currently, there is a growing body of literature that recognises the value of Transformer, especially in the computer vision domain (CV) \cite{ViT, Swin, cluster_vit, cluster_CVPR, ViT2, drop_vit}. When researchers adapt the Transformer from language to vision, the characteristics of the input signal (\textit{i.e.} image) are considered and used as a guide to modify the attention mechanism. In essence, the information contained in a language signal is dense, while that in an image is sparse. Hence, full attention in the standard Transformer is no longer appropriate. As shown in Fig.~\ref{fig_1}(a) and \ref{fig_1}(b), to bridge the gap between language and vision, researchers in vision prefer to use local windows to capture the local information in images. However, using local windows may cause the same object in the image to be separated. To allow connection between different windows, prior works have mainly explored two directions: (1) Shift the attention windows between layers such that the neighboring windows are linked together \cite{Swin}. (2) Cluster the input tokens and perform local attention in feature space \cite{cluster_vit, cluster_CVPR}. For the PSP task, there has also been extensive research regarding the use of Transformer \cite{Monica, ksT, speech_use_trans2, speech_use_trans3, ctnet, Tao2}. Despite the improvements made, little attention has been paid to the characteristic of the input signal (\textit{i.e.} speech), as has been done in computer vision, which is however crucial for extracting paralinguistic information. Therefore, there is an urgent need to exploit the potential of Transformer in PSP by incorporating the features of speech signals. Meanwhile, the standard Transformer is computationally expensive due to its full attention, which makes it difficult to use in practice.

To address these drawbacks, we should rethink the natural structure of speech signals first. As shown in Fig.~\ref{fig_1}(c), we note that a speech signal can be perceived from different perspectives, allowing the information to have different granularities. We refer to this property as the hierarchical relationships of speech. The basic acoustic units that construct the speech signal, from fine to coarse, are frames, phones and words. On the other hand, a word consists of several consecutive phones, and a phone is composed of several consecutive frames. This connection between frames, phones and words is termed the component relationship. Armed with these implicit relationships, we propose a novel attention mechanism for the PSP task, which is illustrated in Fig.~\ref{fig_1}(c). First, we perform, with high computational efficiency, local attention to model the adjacent tokens that belong to the same unit. Statistical durations of phones and words are used to ensure that each local window completely covers the central unit, allowing the boundary information of the central unit to be completely preserved and not separated. The interactions between neighboring units are also considered to comprehensively simulate the inter-unit information. After sufficient learning has been performed in the current stage, we utilize a merging block to aggregate the feature tokens and feed them into the next stage, thus enabling our framework to follow the hierarchical structure of the speech signal. Moreover, to further enrich the information contained in each token, we introduce several learnable word tokens to appropriately incorporate the coarse-grained features into the fine-grained features. The contributions of this paper can be summarized as follows:

\begin{itemize}
\item Based on the component relationship in speech, we propose a unit encoder to capture the intra- and inter-unit information efficiently. To further enhance the extracted features, we utilize a word encoder to effectively integrate the coarse- and fine-grained information.

\item Based on the hierarchical relationship in speech, we construct a hierarchical backbone, called SpeechFormer++, for paralinguistic speech processing. To the best of our knowledge, this is the first study that leverages the characteristics of speech to exploit the potential of Transformer.

\item We evaluate our method on four  benchmark datasets and demonstrate that our SpeechFormer++ substantially outperforms the standard Transformer in terms of performance and computational efficiency. Moreover, SpeechFormer++ achieves superior results compared to the state-of-the-art approaches. Our code is publicly available at \url{https://github.com/HappyColor/SpeechFormer2}.

\end{itemize}

A preliminary version of this work was published in \cite{speechformer}. We have extended our conference version as follows.
In terms of the model structure, (1) we introduce an additional encoder to balance the fine-grained and the coarse-grained information efficiently and achieve superior performance; (2) we investigate the sensitivity of SpeechFormer++ to the statistical durations and offer guidance on applying SpeechFormer++ to various scenarios.
In terms of verifying the effectiveness of the framework, (1) we report results of SpeechFormer++ with pretrained features and hand-crafted features to demonstrate the adaptability of our approach; (2) we compare SpeechFormer++ and other approaches using the same set of features to release the impact of input features; (3) we compare the performances of finetuning the pretrained model directly with several dense layers and learning further deep representation with SpeechFormer++ to prove the importance of our framework; (4) we adopt attention mechanisms from computer vision to demonstrate the necessity of considering the features of speech when modeling speech signal.
In terms of verifying the effectiveness of every module, (1) we conduct a comprehensive ablation study to analyze the indispensability of each module in SpeechFormer++.
In terms of model understanding and interpretation, (1) we visualize the attention weights of Transformer and SpeechFormer++ to determine the reasons behind the improvements; (2) we add more evaluation metrics to each dataset for better justification of results.

The rest of this paper is organized as follows: In Section \uppercase\expandafter{\romannumeral2}, we provide a brief literature review on Transformer and structure-based paralinguistic speech processing. In Section \uppercase\expandafter{\romannumeral3}, we elaborate on the proposed SpeechFormer++ framework. In Section \uppercase\expandafter{\romannumeral4}, we describe the experimental corpora and setup in detail. In Section \uppercase\expandafter{\romannumeral5}, we present our experimental results and analyses. Finally, we draw our conclusions in Section \uppercase\expandafter{\romannumeral6}.

\section{Related Work}
In this section, we systematically introduce the applications of Transformer in different fields and the related research on structure-based paralinguistic speech modeling.

\subsection{Transformer in Language Processing and Computer Vision}
The original Transformer is designed to tackle machine translation tasks in the natural language processing field \cite{Transformer}. As a sequence learning model, Transformer is excellent at modeling the long-range dependencies, while being purely based on the attention mechanism, it dispenses with the recursion and convolution and computes hidden representations in parallel. In general, the raw text signal is first converted into a word embedding sequence via the word and position embedding layers. Then, the output is delivered to a stack of Transformer encoders to produce the final embedding, followed by several Transformer decoders or a task-specific classifier. The Transformer has been applied to various NLP tasks, including question answering \cite{TASLP_QA}, named entity recognition \cite{nlp_use_trans_ner3}, natural language inference \cite{nlp_use_trans_nli1}, semantic textual similarity \cite{TASLP_sentence_embedding} and document classification \cite{nlp_use_trans_dc3}. 

In computer vision, an image is first split into several fixed-size (\textit{e.g.} $16\times16$) patches, followed by a linear projection and a positional embedding layer to yield the input for the Transformer \cite{ViT}. Inspired by CNNs that can be improved by stacking more convolutional layers, researchers have attempted to increase the depth of vision Transformer and solve the attention collapse issue encountered when the model goes deeper \cite{deepvit}. Nevertheless, the self-attention operation scales quadratically with the sequence length, making Transformer computationally expensive and unable to handle numerous tokens in high-resolution images. Consequently, the bulk of the literature has been devoted to enhancing the attention mechanism used to exploit the potential of Transformer in computer vision \cite{Swin, cluster_vit, cluster_CVPR, ViT2, drop_vit}. Typically, Rao \textit{et al.} \cite{drop_vit} evaluated the importance of each token and dropped the useless tokens dynamically and progressively. Liu\textit{ et al.} \cite{Swin} proposed a hierarchical Transformer and performed attention within each shifted window, which greatly reduced the computational cost while also allowing for cross-window interactions. Although this method boosts efficiency, it fails to capture the relationships between distant but similar patches in the image due to the constraint of the window size. To address this limitation, \cite{cluster_vit} first clustered the tokens and then computed self-attention among the related tokens in feature space. 

\subsection{Transformer in Paralinguistic Speech Processing}
There has also been much interest in applying the standard Transformer in the speech domain. Generally, the given raw speech signal is segmented into multiple overlapping frames. Then, the spectral or deep learning-based features are extracted from each frame and used as input for the Transformer \cite{speech_use_trans3, speech_use_trans2, ksT, speech_use_trans1}. In \cite{speech_use_trans2}, stacked multiple Transformer layers were explored to enhance the features extracted for speech emotion recognition. In \cite{speech_use_trans3}, researchers followed the structure of Swin Transformer \cite{Swin} and cut the spectrogram into different patch tokens for sound classification and detection. Although these works demonstrate their effectiveness, they mainly adopt Transformer directly, ignoring the characteristics of speech and task. To overcome this problem, efficient emotion recognition was implemented in \cite{ksT} by utilizing a sparse Transformer to focus more on the emotion-related information. In \cite{Saliency}, an auditory saliency mechanism was studied and applied in a Transformer to adjust the feature embeddings. Additionally, Transformer has been employed for Alzheimer's disease (AD) detection \cite{ad_trans1, ad_trans2} and depression classification \cite{depression_trans1}. For example, Ilias \textit{et al.} \cite{ad_trans1} utilized a pretrained vision Transformer \cite{ViT} to extract acoustic features and achieved remarkable results for dementia detection. Later, Zhu \textit{et al.} \cite{ad_trans2} sought an effective integration of semantic and non-semantic speech information. Both \cite{ad_trans1} and \cite{ad_trans2} encouraged the use of pretrained models. In \cite{depression_trans1}, a Transformer-based network was utilized to extract long-term temporal context information for depression estimation. Based on Transformer, various self-supervised speech representation learning approaches have also been proposed, including wav2vec \cite{wav2vec}, wav2vec 2.0 \cite{wav2vec2} and HuBERT \cite{hubert}. Built on the pretrained self-supervised models, several researches have delivered promising results in the literature \cite{CA-MSER, Monica, reviewer3_1, ksT, ad_trans2, speechformer, sota_w2v2_daic_1, sota_w2v2_meld_1}. Typically, Monica \textit{et al.} \cite{Monica} fine-tuned the pretrained HuBERT model for AD detection and achieved competitive performance. In addition to the paralinguistic tasks, Transformer is also broadly used in speech recognition field \cite{TaoJianhua_asr, reviewer2_2, reviewer2_1}. Typically, Wang \textit{et al.} \cite{reviewer2_2} explored the potential of Transformer-based acoustic models on hybrid speech recognition and achieved significant word error rate improvement over the conventional baselines. Gulati \textit{et al.} \cite{reviewer2_1} novelly proposed a convolution-augmented Transformer, called Conformer, to learn both global interactions and local features effectively. 

\begin{figure*}[th]
\centering
\includegraphics[width=\linewidth]{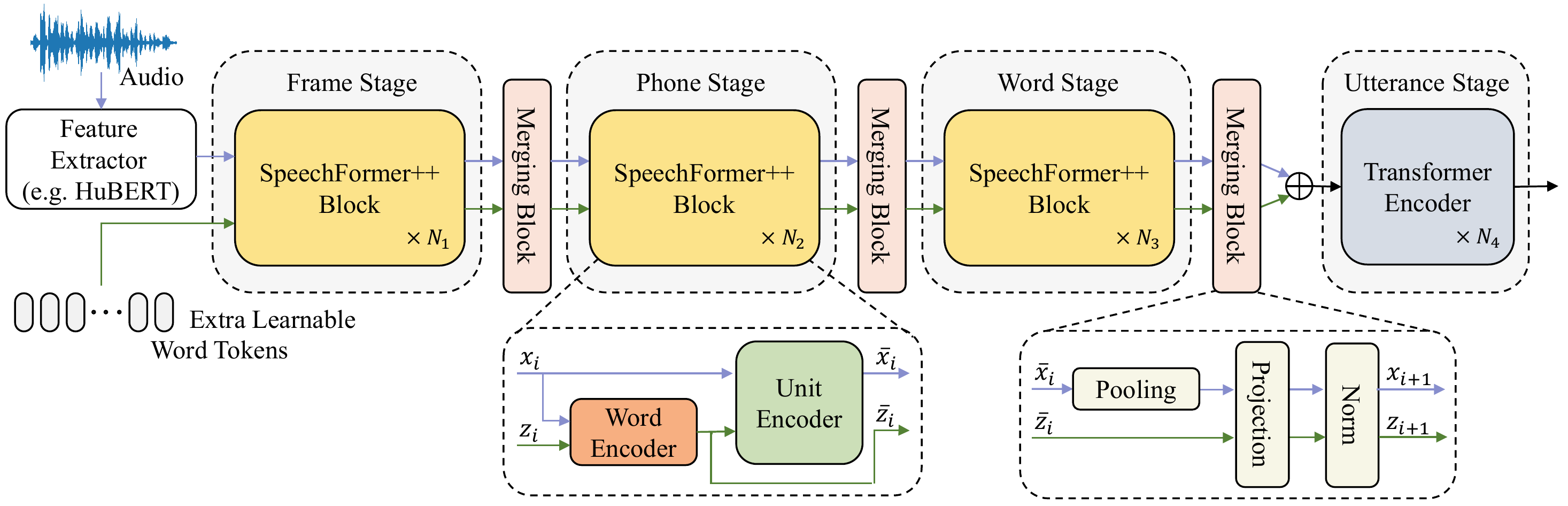}
\caption{Overview of the proposed SpeechFormer++. It mainly consists of four stages and three merging blocks. Raw audio samples and randomly initialized learnable word tokens are used as inputs to our framework. The blue and green lines denote the information flows of the acoustic feature and the extra learnable word tokens, respectively. $\oplus$ represents the concatenation of the tokens for each flow.}
\label{fig_3}
\end{figure*}

\subsection{Structure-Based Paralinguistic Speech Modeling}
The speech signal is structured by different basic units, from fine to coarse, which are frames, phones and words. This natural structure is unique to speech and contains extensive paralinguistic information such as fluency, articulation, prolongation and rhythm. Some previous works have taken advantage of these speech structures to improve system performance \cite{reviewer2_3, reviewer2_4, speech_nature1, speech_nature2, speechformer, speech_nature3}. For example, Zhao \textit{et al.} \cite{reviewer2_3}  trained a hierarchical network for depression severity measurement, where frame-level and sentence-level representations were learned explicitly. However, human speech has more than these two levels. Other levels such as phone-level and word-level can better reflect the pronunciation. To address this problem, a vast majority of works are explored toward hierarchical multi-granularity learning. For example, \cite{reviewer2_4} utilized a hierarchical attention structure with word-level alignment for emotion recognition. In \cite{speech_nature1}, phone- and word-level representations were captured through a GRU network \cite{gru}, using the ground-truth timestamps of every unit. \cite{speech_nature2} aggregated the acoustic embedding for each word based on its corresponding speech frames for information fusion. Although the above methods improve the recognition performance, the requirement of exact timestamps of phones or words in their systems makes them unsuitable for practical applications. Additionally, current studies have not sufficiently explored the hierarchy of speech signal, leaving the full potential of Transformer in the speech domain unexplored. Recently, researchers demonstrated that deep neural networks can integrate the individual basic units across multiple timescales via different integration windows, the sizes of which were yoked to the duration of the units \cite{yoked}. This finding also indicates that the basic units in speech are instructing the model learning. However, existing works on Transformer have not comprehensively considered the audio properties, which is remedied in this paper.

\section{Methodology}
The proposed framework, as shown in Fig~\ref{fig_3}, consists of four stages and three key modules. The unit encoder and word encoder are used for structure-based speech unit learning, and a merging block is employed for structure-based speech unit aggregation. We first clarify the guidelines for model design. Afterwards, we elaborate on the proposed SpeechFormer++.

\begin{figure}[t]
\centering
\includegraphics[width=3.3in]{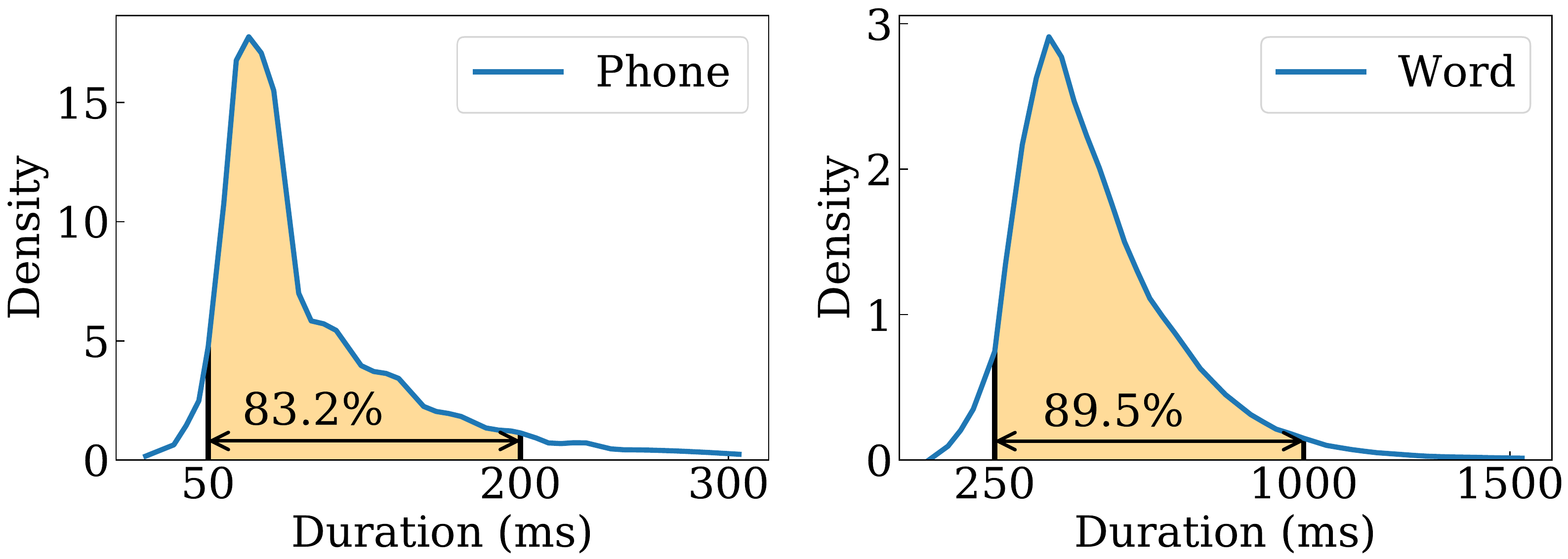}
\caption{Overall duration statistics of phones (left) and words (right) in IEMOCAP, MELD, Pitt and DAIC-WOZ corpora.}
\label{fig_2}
\end{figure}

\subsection{Guiding Principles of Model Design}
\label{statistics}
The statistical duration of the speech unit is the basis for the design of our framework.
Therefore, we first estimate the durations of phones and words on the corpora used in this paper by P2FA \cite{P2FA} toolkit. Since the distribution of the unit duration is similar for each corpus, we illustrate in Fig.~\ref{fig_2} the statistical results obtained by combining all the audio files from four corpora. We note that more than 80\% of phones vary from 50 to 200 ms, and we therefore approximate the shortest and longest durations of phones to be 50 and 200 ms, respectively. Similarly, almost 90\% of words range between 250 to 1000 ms, which we regard as the shortest and longest durations of words. Additionally, we note that the duration of the frame is literally the frame length used when extracting acoustic features, which can be set manually. In addition, the hierarchical pattern in the speech signal aggregates the consecutive units progressively, which sheds new light on the design of the hierarchical framework.

\subsection{Structure-Based Speech Unit Learning}

\subsubsection{Unit Encoder}

Given a speech signal, we first extract its acoustic representations $x_{1} \in \mathbb{R}^{T_1 \times d_1}$, where $T_1$ is the number of frames and $d_1$ is the dimension of each frame embedding. To capture the information about consecutive frames in the frame stage, we employ a unit encoder with window $T_{w1}$ to learn the frame-grained features in $x_1$. Specifically, the frame-grained input feature $x_1$ is split into $T_1$ overlapping segments:
\begin{equation}
\begin{aligned}
\relax[x_{i1}, x_{i2}, \dots, x_{iT_i}] &= OverlapSeg(x_i, T_{wi})  \\ x_{ij} &= x_i\left[j-\tfrac{T_{wi}}{2}:j+\tfrac{T_{wi}}{2}\right]
\label{eq5}
\end{aligned}
\end{equation}
where subscript $i$ denotes the different stages in Fig~\ref{fig_3} (\textit{e.g.,} $i$ = 1 for the frame stage, $i$ = 2 for the phone stage, $i$ = 3 for the word stage and $i$ = 4 for the utterance stage);
$OverlapSeg(\cdot)$ represents the overlapping segmentation, and $j \in [1, T_i]$; $x_i[a:b] \in \mathbb{R}^{(b-a) \times d_i}$ consists of the $a$-th to the $b$-th tokens of $x_i$. The subscript $i$ is equal to 1 because it is currently in the frame stage. Zero padding is employed when the segment is out of range (\textit{e.g.,} when $a < 0 $ or $b > T_i$). The value of $T_{w1}$ is set to the number of tokens that can be contained within 50 ms (the shortest duration of phones) of input $x_1$. Thus, the interactions of nearby frames are learnt. Specifically, the attention in each segment can be written as:
\begin{equation}
\hat{x}_i^{(j)} = Norm(MSA(x_i^{(j)}, x_{ij}, x_{ij}) + x_i^{(j)})
\label{eq6}
\end{equation}
\begin{equation}
\hat{x}_i = Concat(\hat{x}_i^{(1)}, \hat{x}_i^{(2)}, \dots, \hat{x}_i^{(T_i)})
\label{eq7}
\end{equation}
where $x_i^{(j)} \in \mathbb{R}^{1 \times d_i}$ is the $j$-th token in $x_i$, $j \in [1, T_i]$; $\hat{x}_i^{(j)}$ and $\hat{x}_i$ are the updated values of $x_i^{(j)}$ and $x_i$, respectively. $MSA(Q,K,V)$ represents the Multi-Head Self-Attention (MSA) mechanism with inputs query $Q$, key $K$ and value $V$. More details of MSA can be found in \cite{Transformer}. $Norm(\cdot)$ represents the layer normalization \cite{LN} throughout the paper. When performing attention, the query $x_i^{(j)}$ denotes the central feature token in the current overlapping segment. 

In the phone stage, we assume the phone-grained input feature to be $x_2 \in \mathbb{R}^{T_2 \times d_2}$, where $T_2$ denotes the number of phone tokens and $d_2$ denotes the dimensions of phone embeddings. Each token contained in $x_2$ is the representation of a phone or subphoneme, which is produced by the merging block using the output of the frame stage (described in \ref{MB}) and fed into the phone stage. To model a phone and the interactions with its neighbors, the value of window $T_{w2}$ is set to the number of tokens that can be contained within 400 ms (twice the longest duration of phones) of $x_2$. Thus, each segment covers consecutive phones, and the central phone is unbroken. Finally, the attention calculation in the phone stage follows Eqs.~\ref{eq5}-\ref{eq7} with $i=2$.

\begin{figure}[t]
\centering
\includegraphics[width=0.8\linewidth]{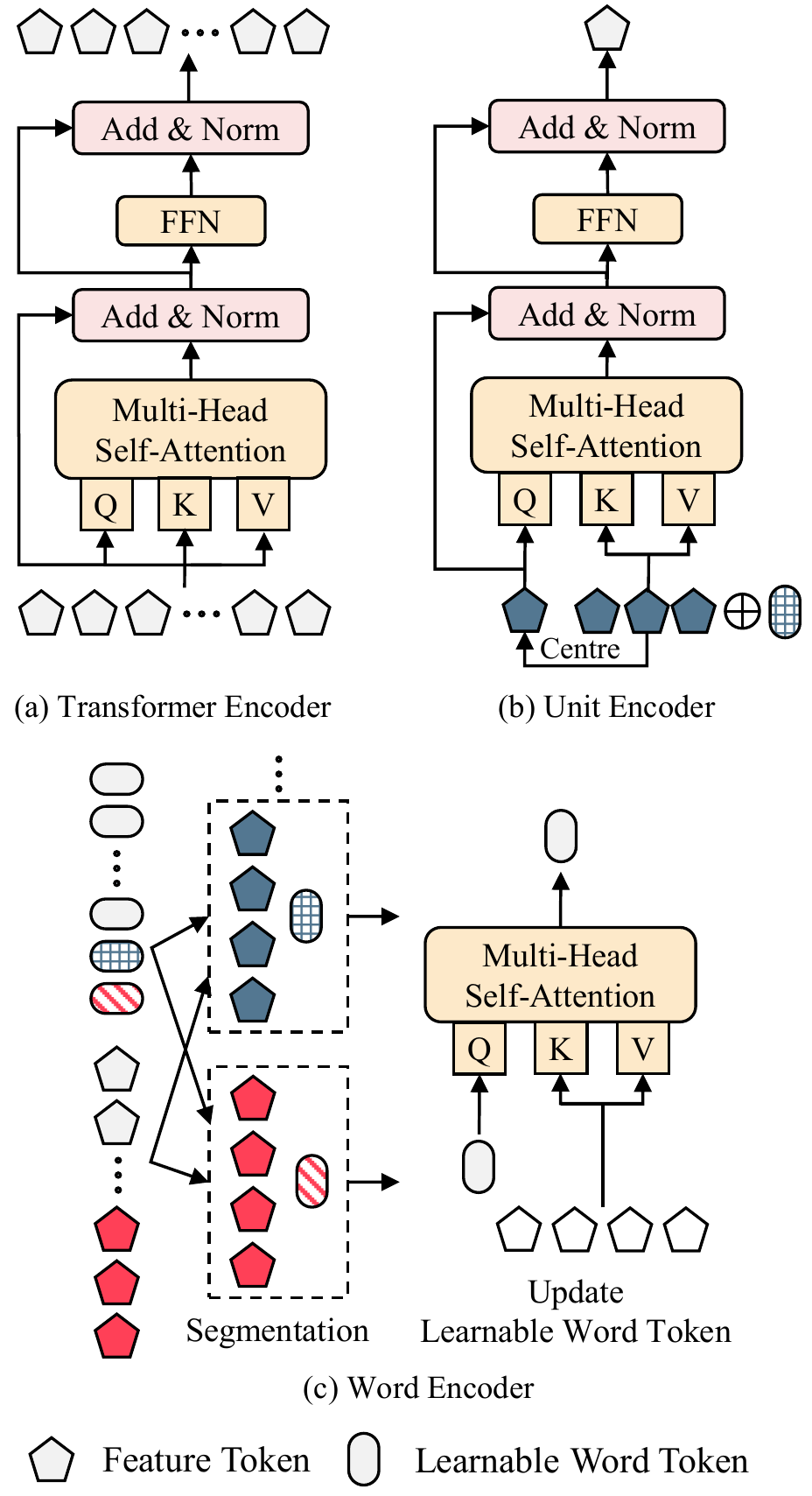}
\caption{Architectures of the (a) Transformer encoder, (b) unit encoder and (c) word encoder. Feature token denotes the acoustic feature.}
\label{fig_5}
\end{figure}

Similarly, in the word stage, its word-grained input feature is $x_3 \in \mathbb{R}^{T_3 \times d_3}$, where $T_3$ and $d_3$ represent the number of word tokens and the dimensions of word embeddings, respectively. It is produced by a merging block using the output of the phone stage (described in \ref{MB}). To capture the intra- and inter-word information, the window size $T_{w3}$ in the word stage is set to the number of tokens that can be contained within 2000 ms (twice the longest duration of words) of $x_3$. The attention mechanism is then invoked in the overlapping segments, each of which contains a central word and its surrounding context. The computational process follows Eqs.~\ref{eq5}-\ref{eq7} with $i=3$. 

\subsubsection{Word Encoder}
The proposed unit encoder is able to model the fine-grained features efficiently. However, its receptive field is limited by the size of the attention window. To take the coarse-grained information into account, we  propose a word encoder (Fig.~\ref{fig_5}(c)) to inject the coarse-gained information into each unit encoder. We first create several learnable word tokens $z_1 \in \mathbb{R}^{T_z \times d_1}$ for the frame stage, where $T_z$ indicates the approximate number of words in the utterance. Concretely, the value of $T_z$ is equal to the total duration of the utterance divided by 1000 ms (the longest duration of words). $z_2 \in \mathbb{R}^{T_z \times d_2}$ for the phone stage and $z_3 \in \mathbb{R}^{T_z \times d_3}$ for the word stage are produced by the merging block (described in \ref{MB}). Then, the input $x_i$ is evenly grouped into $T_z$ non-overlapping segments. Each learnable word token is required to learn the coarse-grained features about the corresponding segment. The learning process is  as follows:
\begin{equation}
\begin{aligned}
\relax [s_{i1}, s_{i2}, \dots, s_{iT_z}] &= EvenSeg\left(x_i, \tfrac{T_i}{T_z}\right) \\
\ s_{ij} &= x_i\left[\tfrac{T_i}{T_z} \times (j-1):\tfrac{T_i}{T_z} \times j\right]
\label{eq10}
\end{aligned}
\end{equation}
\begin{equation}
\ \ \ \ \; \bar{z}_i^{(j)} = MSA(z_i^{(j)}, s_{ij}, s_{ij})
\label{eq11}
\end{equation}
where $EvenSeg(\cdot)$ denotes the non-overlapping segmentation, $s_{ij}$ is the $j$-th non-overlapping segment of $x_i$ and $j \in [1, T_z]$, $z_i^{(j)}$ denotes the $j$-th learnable word token in $z_i$ and $\bar{z}_i^{(j)}$ is the updated value of $z_i^{(j)}$. Since the interactions between words are modeled by the word stage of SpeechFormer++, we perform non-overlapping segmentation in the word encoder. Note that the number of non-overlapping segments is always identical to that of the learnable word tokens and remains constant across different stages.

Then, we pass the learnt $\bar{z}_i$ to each unit encoder in the $i$-th stage, allowing the unit encoders to take the coarse-grained information into consideration while modeling locally. As shown in Fig.~\ref{fig_5}(b), each acoustic segment is enhanced by its corresponding learnable word token, which is then fed into the MSA and FFN layers. The complete calculation flow in the unit encoder (Fig.~\ref{fig_5}(b)) is  as follows:
\begin{equation}
e_i^{jk} = Concat(\bar{z}_i^{(k)}, x_{ij}),\, \  k=Ceil\left[\tfrac{j\times T_z}{T_i}\right]
\label{eq12}
\end{equation}
\begin{equation}
\hat{x}_i^{(j)} = Norm(MSA(x_i^{(j)}, e_i^{jk}, e_i^{jk}) + x_i^{(j)})
\label{eq13}
\end{equation}
\begin{equation}
\bar{x}_i = Norm(FFN(\hat{x}_i)+\hat{x}_i)
\label{eq14}
\end{equation}
where $e_i^{jk} \in \mathbb{R}^{(1+T_{wi}) \times d_i}$ is the enhanced segment, $Ceil[\cdot]$ rounds a number upward to its nearest integer and $j \in [1, T_i]$; $FFN(\cdot)$ denotes the feed-forward network and $\bar{x}_i$ denotes the final output of the unit encoder. 
The parameters in the MSA are shared between the unit encoder and the word encoder, keeping the size of the model unchanged.
In addition, a unit encoder and a word encoder constitute a basic SpeechFormer++ block. Multiple SpeechFormer++ blocks are stacked to form a stage in our proposed framework.

\subsection{Structure-Based Speech Unit Aggregation}
\label{MB}
\subsubsection{Merging Block}
Inspired by the hierarchical property of speech signals that can be progressively categorized into frames, phones and words, we propose a merging block to generate the relevant features under the instruction of the statistical durations of the speech units. As shown in Fig.~\ref{fig_3},  merging blocks are used between each stage. Initially, the acoustic input of the frame stage $x_1$ represents the features of each frame from the original speech signal. To provide the phone-grained input to the phone stage, we apply average pooling over the output of the frame stage $\bar{x}_1$ with a merging scale $M_1$ of 50 ms (the shortest duration of phones). Then, a linear projection and layer normalization are performed to create the phone-grained feature $x_2$. The information contained every 50 ms is aggregated into a token in $x_2$ such that each token in $x_2$ represents the information of a subphoneme. Analogously, the merging scale $M_2$ is set to 250 ms (the shortest duration of words) when attempting to generate the word-grained input $x_3$ for the word stage, making each token in $x_3$ a representation of a subword. Finally, the last merging block is applied to the output of the word stage $\bar{x}_3$ while merging scale $M_3$ is set to 1000 ms (the longest duration of words) to roughly simulate the number of words in the utterance sample. 
The learnable word tokens $z_i$ represent the coarse-grained features in words, and thus, we do not have to aggregate them.
Formally, the merging block is defined as:
\begin{align}
    x_{i+1} &=Norm(AvgPool(\bar{x}_i, M_i)W_i + b_i) \\
    z_{i+1} &= Norm(\bar{z}_iW_i + b_i)
    \label{eq16}
\end{align}
where $AvgPool(x, M)$ represents an average pooling layer performed on $x$ with window size and stride equal to $M$; $W_i \in \mathbb{R}^{d_i \times d_{i+1}}$ and $b \in \mathbb{R}^{d_{i+1}}$ are to be learned parameters; $\bar{x}_i$ and $\bar{z}_i$ denote the outputs of the $i$-th stage and $x_{i+1}$ and $z_{i+1}$ denote the inputs of the next stage, $i \in \{1, 2, 3\}$.

\begin{figure}[t]
\centering
\includegraphics[width=\linewidth]{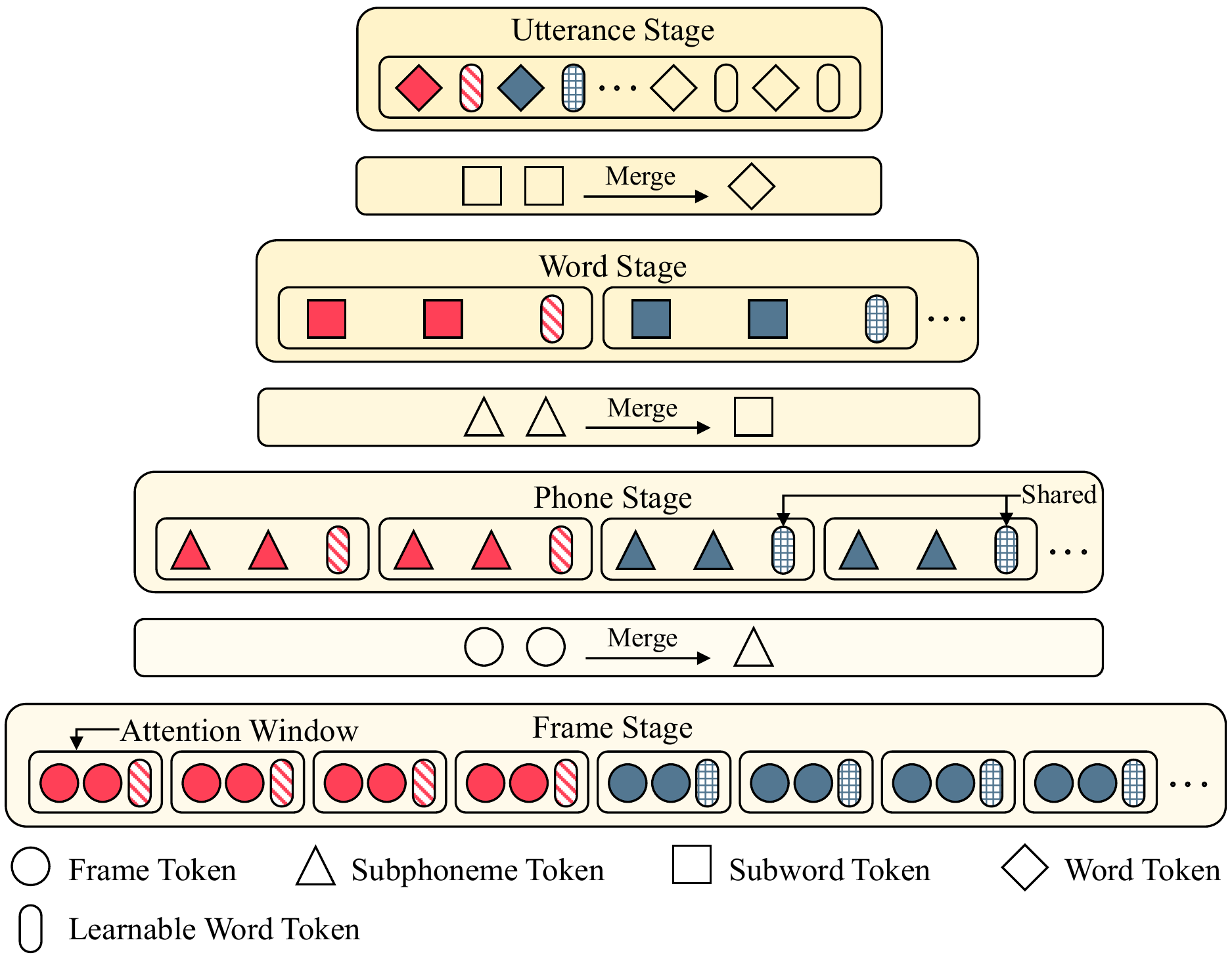}
\caption{Overview of the calculation process in SpeechFormer++. In practice, the segmentation is overlapping across consecutive windows. The learnable word tokens of the same color are shared in the same stage. We show the case when window size and merging scale are 2 in each stage for brevity.
}
\label{fig_4}
\end{figure}

The outputs of the third merging block are concatenated together and fed into the utterance stage, which is a stack of standard Transformer encoders, to model the speech signal globally. 
The overview of the computational flow in our proposed method is illustrated in Fig.~\ref{fig_4}. The acoustic tokens are aggregated progressively to imitate the structural pattern in the speech signal, and the attention is guided by the characteristics of speech. The final output of the utterance stage is pooled in the temporal dimension and is passed to a classifier, which is composed of two linear projections with an activation function in between, to yield the final classification results.

\subsection{Loss Function}
We choose the categorical cross-entropy loss (CCE) as the objective function in this paper. Suppose we have $S$ samples and $C$ possible categories. The CCE can be represented as:
\begin{equation}
    CCE = -\frac{1}{S}\sum_{s=1}^S\sum_{c=1}^C y_{sc}\log_2(\hat{y}_{sc})
    \label{eq19}
\end{equation}
where $\hat{y}_{sc} \in \mathbb{R}^{1}$ denotes the predicted probability that the $s$-th sample belongs to class $c$ and $y_{sc} \in \mathbb{R}^{1}$ is 1 when c is equal to the ground-truth label and 0 otherwise.

\subsection{Complexity Analysis}
Supposing inputs $x \in \mathbb{R}^{T \times d}$, $z \in \mathbb{R}^{T_z \times d}$ and window size is $T_w$. The computational complexities of the MSA in Transformer and SpeechFormer++ (S-MSA) are:
\begin{equation}
    \Omega(MSA) = 4Td^2+2T^2d
\label{eq17}
\end{equation}
\begin{equation}
    \Omega(S\mbox{-}MSA) = 4(T+T_z)d^2+2T(T_w+2)d
\label{eq18}
\end{equation}
Note that we omit softmax computation in determining complexity. When $T_w$ and $T_z$ are fixed, $\Omega(S\mbox{-}MSA)$ scales linearly with the sequence length $T$, while $\Omega(MSA)$ in the standard Transformer scales quadratically. Moreover, the values of $T_w$ and $T_z$ are much smaller than that of $T$ in practice. When features go through a merge block, the number of tokens is greatly reduced, enabling the computational cost of the later layers in SpeechFormer++ to become fairly low. The cost of the merging block is negligible compared to the total complexity and the model size.

\section{Experimental Setup}

\subsection{Datasets and Evaluation Metric}
IEMOCAP \cite{IEMOCAP} is the most commonly used dataset in the speech emotion recognition field. It contains 12 hours of audio data and consists of five sessions, each of which has one male speaker and one female speaker. 5,531 utterances from four emotion categories: angry, neutral, happy\footnote{We merge the excited samples with the happy samples in IEMOCAP.} and sad, are considered in this work. To train and test the model, we conduct experiments in the leave-one-session-out cross-validation strategy. Specifically, samples from 4 sessions are used for training, and the remaining session is regarded as the testing set, which is repeated 5 times until all different sessions are used for training and testing. We evaluate at each epoch the model on the testing set and the reported results are the average scores of the 5-fold experiments.

MELD \cite{meld} is the second dataset we used for emotion recognition. The dataset contains 13,708 utterances from the \textit{Friends} TV series, divided into 7 emotion classes: anger, disgust, sadness, joy, neutral, surprise and fear. Since this dataset has been officially divided into training, validation and testing sets, we use the validation set for hyperparameter turning. The model with the best performance on the validation set across epochs is evaluated on the testing set. Finally, the results on the testing set are reported.

Pitt \cite{pitt} is a classical dataset used in the AD detection field. To produce the narrative speech recordings, the AD patients and healthy controls are asked to take the ``Cookie Theft" picture description task from the Boston Diagnostic Aphasia Examination \cite{cookie}. To evaluate the model on Pitt dataset, the speaker-independent 10-fold cross-validation technique is implemented. Similar to IEMOCAP, we evaluate at each epoch the model on the testing set and the reported results are the average scores of the 10-fold experiments

DAIC-WOZ \cite{avec}, used in AVEC 2017 \cite{avec2017}, is a subset of the Distress Analysis Interview Corpus (DAIC) \cite{avec}. This dataset contains training, validation and testing sets originally, and a label depressed/not depressed is assigned to each clinical interview recording in the training and validation sets, but the labels of the test data are not provided. Therefore, we randomly select 20\% of the training data for hyperparameter turning and checkpoint selection. Finally, the results on its original validation set are reported.

Following previous works\cite{ctnet, Monica}, we apply four evaluation metrics to evaluate the performance of different learning algorithms: weighted accuracy (WA), unweighted accuracy (UA), weighted average F1 (WF1) and macro average F1 (MF1). The above criteria can be formulated as:
\begin{align}
    WA &= \frac{1}{\sum_{c=1}^CS_c}\sum_{c=1}^C S_c\times Acc(c)\\
    UA &= \frac{1}{C}\sum_{c=1}^C Acc(c)\\
    WF1 &= \frac{1}{\sum_{c=1}^CS_c}\sum_{c=1}^C S_c\times F1(c)\\
    MF1 &= \frac{1}{C}\sum_{c=1}^C F1(c)
    \label{eq23}
\end{align}
where $S_c$ denotes the number of samples of the $c$-th category and $Acc(c)$ and $F1(c)$ are the classification accuracy and F1 score of the $c$-th category, respectively.

For speech emotion recognition on IEMOCAP and MELD, we aim to predict the discrete emotion labels for each individual utterance. While conducting neurocognitive disorder analyses (\textit{i.e.}, namely, Alzheimer's disease detection on Pitt and depression classification on DAIC-WOZ), we first receive a dialogue, and we then crop out the utterances of the participant based on the provided transcription timestamps. Subsequently, the utterances are processed and predicted, and a majority vote is applied to yield a subject-level prediction, which is used for final evaluation.

\subsection{Implementation Details}
\subsubsection{Acoustic Features} Encouraged by the success of self-supervised learning models in various speech tasks, we utilize the pretrained HuBERT-large \cite{hubert} model to extract the acoustic features. Specifically, the duration of each frame processed in HuBERT is 25 ms, and the hop length used when yielding the overlapping frames is 20 ms. The overlap between consecutive frames is 5 ms. In total, 1024-dimensional frame-grained features are extracted for each utterance sample. Recently, it has been reported that the output from the middle layer has the most pronunciation-related features \cite{12layer}. Hence, we use the output from the 12-th layer of the 24-layer Transformer encoder in HuBERT. Unless otherwise stated, the pretrained self-supervised models are only used to extract the acoustic features and will not be involved in the training procedure. The max sequence lengths are set to 326, 224, 328 and 426 for IEMOCAP, MELD, Pitt and DAIC-WOZ, respectively, because 80\% of samples in each dataset are shorter than the corresponding set sequence lengths. We also report the results of SpeechFormer++ with hand-crafted features, such as 80-dimensional log-mel filter bank coefficients (FBANK). Unless otherwise stated, Hubert features are used in SpeechFormer++.

\begin{table}[t]
    \caption{Performance and computational efficiency of Transformer and SpeechFormer++ using HuBERT features on IEMOCAP. Gain indicates the relative improvement (+) or reduction (-)}
    \label{tab_1}
    \centering
    \setlength{\tabcolsep}{4pt}{
    \begin{tabular}{c||cc||ccc}
    \hline
    Method         & Params    & FLOPs     & WA       & UA     & WF1     \\ \hline
    Transformer    & 63.64M    & 23.12G    & 0.685    & 0.701  & 0.692 \\ 
    SpeechFormer++ & 66.79M    & 6.55G     & \textbf{0.705}   & \textbf{0.715} & \textbf{0.707}  \\ \hline
    Gain           & +4.95\%   & -71.67\%  & +2.92\%  & +2.00\%  & +2.17\% \\ \hline
    \end{tabular}}
\end{table}

\begin{table}[t]
    \caption{Comparison with state-of-the-art methods on IEMOCAP. All systems apply audio as input for a fair and direct comparison. h/c=hand-crafted, w2v2=wav2vec 2.0}
    \label{tab_2}
    \centering
    \begin{threeparttable}
    \begin{tabular}{ccc||ccc}
    \hline
    Method   & Features  & Year & WA  & UA & WF1   \\ \hline
    STC\cite{stc}       &  H/C     & 2021       & 0.613   & 0.604 & 0.617  \\
    \tnote{$\dagger$}\; ISNet\cite{ISNet}     &  H/C   & 2022         & 0.704   & 0.650  & - \\
    LSTM-GIN\cite{LSTM-GIN}  &  H/C & 2021   & 0.647   & 0.655  & -\\
    SUPERB\cite{superb} & w2v2  & 2021 & 0.656 & - & -  \\ 
    SUPERB\cite{superb} & HuBERT  & 2021 & 0.676 & - & -  \\ 
    CA-MSER\cite{CA-MSER}  & H/C + w2v2   & 2022  & 0.698   & 0.711  & - \\ \hline
    SpeechFormer++   & H/C  & 2022  & 0.645   & 0.658 & 0.649 \\
    SpeechFormer++   & HuBERT  & 2022  & \textbf{0.705}   & \textbf{0.715} & \textbf{0.707} \\ \hline
    \end{tabular}
    \begin{tablenotes}
        \footnotesize
        \item[$\dagger$] Speaker information is used.
    \end{tablenotes}
    \end{threeparttable}
\end{table}

\subsubsection{Training Details} We train SpeechFormer++ in an end-to-end manner using a Nvidia GeForce RTX 2080 Ti GPU. The total number of training epochs are set to 120, 120, 80 and 60 for IEMOCAP, MELD, Pitt and DAIC-WOZ, respectively, and their initial learning rates are set to 0.0005, 0.0005, 0.001 and 0.0001, respectively. The learning rate gradually drops to 1\% of the original by cosine annealing. The batch size is set to 32. The model is updated by SGD with momentum 0.9. The number of attention heads used in MAS is set to 8. For the sake of simplicity, the dimensions of $x_i$ and $z_i$, $i \in \{1,2,3,4\}$, are all set to 1024. Unless otherwise stated, the number of layers employed in the frame stage $N_1$, phone stage $N_2$ and word stage $N_3$ are 2, 2 and 4, respectively, and the number of Transformer encoders used in the utterance stage $N_4$ is 4. As a result, the total number of layers of SpeechFormer++ is 12.

\section{Results and Discussion}
In this section, we report the experiments conducted on four corpora, including three recognition tasks. First, we compare the proposed SpeechFormer++ with the standard Transformer architecture in terms of performance and computational efficiency. To be consistent with the settings of SpeechFormer++, a total of 12 Transformer encoders are used in the standard Transformer framework. Second, we present a comparison with previous works. Finally, we perform extensive ablation studies to better understand the effectiveness of each module.

\begin{table}[t]
    \caption{Performance and computational efficiency of Transformer and SpeechFormer++ using HuBERT features on MELD.\\Gain indicates the relative improvement (+) or reduction (-)}
    \label{tab_3}
    \centering
    \setlength{\tabcolsep}{4pt}{
    \begin{tabular}{c||cc||ccc}
    \hline
    Method  & Params  & FLOPs & WA & UA & WF1 \\ \hline
    Transformer  & 63.64M    & 15.33G  & 0.485 & 0.257  & 0.454     \\ 
    SpeechFormer++ & 66.79M    & 4.51G   & \textbf{0.510} & \textbf{0.273}  & \textbf{0.470}     \\ \hline
    Gain           & +4.95\%   & -70.58\% & +5.16\% & +6.23\% & +3.52\%   \\ \hline
    \end{tabular}}
\end{table}

\begin{table}[t]
    \caption{Comparison with state-of-the-art methods on MELD. All systems apply audio as input for a fair and direct comparison. h/c=hand-crafted, w2v2=wav2vec 2.0}
    \label{tab_4}
    \centering
    \begin{threeparttable}
    \begin{tabular}{ccc||ccc}
    \hline
    Method   & Features  & Year  & WA & UA & WF1        \\ \hline
    \tnote{$\dagger$}\; ConGCN\cite{ConGCN} & H/C & 2019 & - & - & 0.422    \\
    MMFA-RNN\cite{MMFA-RNN} & H/C & 2020 & 0.488  & - & 0.423    \\
    \tnote{$\dagger$}\; MM-DFN\cite{MM-DFN} & H/C & 2022 & -  & - & 0.427    \\
    \tnote{$\dagger$}\; CTNet\cite{ctnet} & H/C & 2021  & 0.469 & - & 0.382 \\
    Sharma\cite{sota_w2v2_meld_1} & w2v2 & 2022 & 0.498 & - & - \\ \hline
    SpeechFormer++  & H/C  & 2022 & 0.480 & 0.236 & 0.429   \\
    SpeechFormer++  & HuBERT  & 2022 & \textbf{0.510} & \textbf{0.273} & \textbf{0.470}    \\ \hline
    \end{tabular}
    \begin{tablenotes}
        \footnotesize
        \item[$\dagger$] Conversation context and speaker information are used.
    \end{tablenotes}
    \end{threeparttable}
\end{table}

\begin{table}[t]
    \caption{Performance and computational efficiency of Transformer and SpeechFormer++ using HuBERT features on Pitt.\\Gain indicates the relative improvement (+) or reduction (-)}
    \label{tab_5}
    \centering
    \setlength{\tabcolsep}{4pt}{
    \begin{tabular}{c||cc||ccc}
    \hline
    Method         & Params     & FLOPs       & WA       & UA    & WF1   \\ \hline
    Transformer    & 63.64M     & 23.28G      & 0.789    & 0.790  & 0.780 \\ 
    SpeechFormer++ & 66.79M     & 6.58G       & \textbf{0.813}    & \textbf{0.816}  & \textbf{0.808} \\ \hline
    Gain           & +4.95\%    & -71.74\%    & +3.04\%  & +3.29\% & +3.59\% \\ \hline
    \end{tabular}}
\end{table}

\begin{table}[t]
    \caption{Comparison with state-of-the-art methods on Pitt. All systems apply audio as input for a fair and direct comparison. h/c=hand-crafted, w2v2=wav2vec 2.0}
    \label{tab_6}
    \centering
    \begin{threeparttable}
    \begin{tabular}{ccc||ccc}
    \hline
    Method  & Features  & Year   & WA  & UA & WF1    \\ \hline
    GCNN\cite{GCNN} & H/C & 2018 & 0.736 & - & - \\
    Makiuchi\cite{Makiuchi}  & H/C & 2021 & 0.731  & 0.731 & 0.732 \\
    Autoencoder\cite{Autoencoder}  & H/C    & 2022   & 0.739   & 0.641 & 0.621 \\
    P\'{e}rez-Toro\cite{pitt_use_w2v2} & w2v2 & 2022 & - & - & 0.720 \\
    Monica\cite{Monica} & HuBERT & 2022 & 0.740 & 0.740 & 0.745 \\ \hline
    SpeechFormer++   & H/C  & 2022  & 0.742   & 0.738  & 0.727 \\
    SpeechFormer++   & HuBERT  & 2022  & \textbf{0.813}   & \textbf{0.816}  & \textbf{0.808} \\ \hline
    \end{tabular}
    \end{threeparttable}
\end{table}

\begin{table}[t]
    \caption{Performance and computational efficiency of Transformer and SpeechFormer++ using HuBERT features on DAIC-WOZ.\\Gain indicates the relative improvement (+) or reduction (-)}
    \label{tab_7}
    \centering
    \setlength{\tabcolsep}{4pt}{
    \begin{tabular}{c||cc||ccc}
    \hline
    Method         & Params   & FLOPs  & WA & UA & MF1       \\ \hline
    Transformer    & 63.64M   & 31.26G & 0.686 & 0.661    & 0.658     \\ 
    SpeechFormer++ & 66.79M   & 8.53G  & \textbf{0.771} & \textbf{0.726}    & \textbf{0.709}     \\ \hline
    Gain           & +4.95\%    & -72.71\%    & +12.39\% & +9.83\% & +7.75\%   \\ \hline
    \end{tabular}}
\end{table}

\begin{table}[t]
    \caption{Comparison with state-of-the-art methods on DAIC-WOZ. All systems apply audio as input for a fair and direct comparison. h/c=hand-crafted, w2v2=wav2vec 2.0}
    \label{tab_8}
    \centering
    \begin{threeparttable}
    \begin{tabular}{ccc||ccc}
    \hline
    Method   & Features  & Year & WA  & UA  & MF1        \\ \hline
    FVTC-CNN\cite{FVTC-CNN}  &  H/C  & 2020 & 0.735 & 0.656 & 0.640 \\
    EmoAudioNet\cite{EmoAudioNet} &  H/C & 2021 & 0.732 & 0.649 & 0.653 \\
    Saidi\cite{Saidi} &  H/C & 2020  & 0.680 & 0.680  & 0.680 \\
    Solieman\cite{Solieman} &  H/C & 2021 & 0.660 & 0.615 & 0.610 \\ 
    SIMSIAM-S\cite{daic_use_hubert} & HuBERT & 2022 & 0.703 & - & -  \\
    TOAT\cite{sota_w2v2_daic_1} & w2v2 & 2022 & 0.717 & 0.429 & 0.480 \\ \hline
    SpeechFormer++ & H/C  & 2022 & 0.743 & \textbf{0.754} & \textbf{0.733} \\
    SpeechFormer++ & HuBERT  & 2022 & \textbf{0.771} & 0.726 & 0.709 \\ \hline
    \end{tabular}
    \end{threeparttable}
\end{table}

\subsection{Speech Emotion Recognition on IEMOCAP}
\subsubsection{Comparison to Transformer} Table~\ref{tab_1} presents the results of Transformer and the proposed SpeechFormer++ on IEMOCAP. Our SpeechFormer++ has a slightly larger model size due to the addition of the merging blocks. However, the theoretical computational complexity (FLOPs) of SpeechFormer++ is greatly reduced (by 71.67\%) compared to Transformer. Meanwhile, our model boosts performance consistently (0.705 vs. 0.685 in WA, 0.715 vs. 0.701 in UA and 0.707 vs. 0.692 in WF1), meaning that our model is efficient and effective.

\subsubsection{Comparison to Previous State-of-the-Art} Table~\ref{tab_2} lists the results of SpeechFormer++ and existing works on IEMOCAP. Our model with HuBERT fetures achieves 0.705 WA, 0.715 UA and 0.707 WF1, surpassing the previous best results. Our SpeechFormer++ with hand-crafted features outperforms STC \cite{stc} (0.645 vs. 0.613 in WA, 0.658 vs. 0.604 in UA and 0.649 vs. 0.617 in WF1) and achieves comparable results to LSTM-GIN \cite{LSTM-GIN} (0.645 vs. 0.647 in WA and 0.658 vs. 0.655 in UA) under the same experimental setup. SpeechFormer++ obtains inferior results compared to ISNet \cite{ISNet}. We suspect this is because ISNet is equipped with a carefully designed individual benchmark to alleviate the problem of interindividual emotion confusion. In addition, speaker information is used in ISNet. Our SpeechFormer++ is a general backbone and can be employed in ISNet for further improvement.

\subsection{Speech Emotion Recognition on MELD}
\subsubsection{Comparison to Transformer} The results on MELD are shown in Table~\ref{tab_3}. Compared to the standard Transformer, SpeechFormer++ yields a relative improvement of 5.16\% in WA, 6.23\% in UA and 3.52\% in WF1. Although the model size of SpeechFormer++ is slightly larger, the computational effort of SpeechFormer++ is reduced from 15.33G to 4.51G, which is a 70.58\% relative reduction.

\subsubsection{Comparison to Previous State-of-the-Art} Table~\ref{tab_4} compares SpeechFormer++ with previous state-of-the-art models on MELD. It can be observed that SpeechFormer++ with HuBERT features noticeably outperforms the previous works by a large margin of +3.1\% WF1 and +1.2\% WA. When compared under the hand-crafted features, SpeechFormer++ outperforms ConGCN \cite{ConGCN}, MMFA-RNN \cite{MMFA-RNN}, MM-DFN \cite{MM-DFN} and CTNet \cite{ctnet} in terms of WF1. Note that SpeechFormer++ is simply applied in MELD and does not utilize the context and speaker information. This demonstrates the potential of SpeechFormer++ and the possibility of further improvement.

\subsection{Alzheimer's Disease Detection on Pitt}
\subsubsection{Comparison to Transformer} As shown in Table~\ref{tab_5}, our SpeechFormer++ once again beats the standard Transformer framework on Pitt in terms of WA and UA while having much lower FLOPs. In detail, the results of SpeechFormer++ are +2.4 WA, +2.6 UA and +2.8 WF1 superior to the Transformer with a comparable model size (66.79M vs. 63.64M) and a significantly lower computational burden (6.58G vs. 23.28G).

\subsubsection{Comparison to Previous State-of-the-Art} Table~\ref{tab_6} gives the comparison among SpeechFormer++ with existing works on Pitt. Our method using HuBERT features outperforms other comparisons with promising gains: +7.7\% WA over \cite{GCNN}, +8.2\% WA over \cite{Makiuchi}, +8.8\% WF1 over \cite{pitt_use_w2v2}, +7.4\% (+17.5\%) WA (UA) over \cite{Autoencoder} and +7.3\% (+7.6\%) WA (UA) over \cite{Monica}. Also, our method using FBANK outperforms the competitors using hand-crafted features in terms of WA and UA.

\subsection{Depression Classification on DAIC-WOZ}

\subsubsection{Comparison to Transformer} Results of the standard Transformer and SpeechFormer++ on DAIC-WOZ corpus are shown in Table~\ref{tab_7}. For Transformer, the FLOPs reach 31.26G since the durations of audio samples in DAIC-WOZ are overall longer than those of the other three corpora. The computation effort grows rapidly as the length of the input sequence increases. Not surprisingly, our SpeechFormer++ delivers superior performance (0.771 vs. 0.686 in WA, 0.726 vs. 0.661 in UA and 0.709 vs. 0.658 in MF1) while keeping the FLOPs at a relatively low level (8.53G)

\subsubsection{Comparison to Previous State-of-the-Art} The comparison results of the proposed SpeechFormer++ and the previous works on DAIC-WOZ are presented in Table~\ref{tab_8}. Our method with HuBERT features outperforms currently advanced approaches by a considerable margin in all metrics. Additionally, SpeechFormer++ using FBANK features achieves state-of-the-art compared to other hand-crafted feature-based methods, drawing the improvement of 0.3\%$\sim$15.7\% on WA, 7.4\%$\sim$16.8\% on UA and 5.3\%$\sim$18.1\% on MF1.

\subsection{Comparison Under the HuBERT Features}
To release the impact of input features, we compare SpeechFormer++ with other approaches using HuBERT features. Experimental results in Table~\ref{tab_hubert} demonstrate that SpeechFormer++ shows superior performance on four corpora. For IEMOCAP, our method shows an absolute improvement of 2.9\%$\sim$4.8\% on WA, 4.6\%$\sim$5.6\% on UA and 4.5\%$\sim$6.4\% on WF1 over other competitors. For MELD, our method shows an absolute improvement of 1.0\%$\sim$3.3\% on WA, 1.9\%$\sim$3.0\% on UA and 1.2\%$\sim$2.4\% on WF1 over other competitors. For Pitt, our method shows an absolute improvement of 1.9\%$\sim$7.3\% on WA, 2.5\%$\sim$7.6\% on UA and 3.4\%$\sim$6.3\% on WF1 over other competitors. For DAIC-WOZ, our method shows an absolute improvement of 5.7\%$\sim$8.5\% on WA, 2.3\%$\sim$2.5\% on UA and 1.5\%$\sim$3.3\% on MF1 over other competitors. The reason lies in that other methods ignore the structural features of the speech signal, which is remedied in SpeechFormer++. These results verify the effectiveness of the proposed method.

\begin{table}[t]
    \caption{Performance of different approaches using HuBERT features. F1 stands for MF1 on DAIC-WOZ, and WF1 on other datasets.
    }
    \label{tab_hubert}
    \centering
    \begin{threeparttable}
    \begin{tabular}{c||c||ccc}
    \hline
    Dataset                   & Method  & WA   & UA  & F1   \\ \hline
    \multirow{4}{*}{IEMOCAP}  & SUPERB\cite{superb} & 0.676 & - & -  \\
                              & \tnote{$\dagger$}\; STC\cite{stc} & 0.657	& 0.659	& 0.643  \\
                              & \tnote{$\dagger$}\; LSTM-GIN\cite{LSTM-GIN} & 0.661 & 0.669 & 0.662 \\ \cline{2-5}
                              & SpeechFormer++ & \textbf{0.705} & \textbf{0.715} & \textbf{0.707}  \\ \hline
    \multirow{3}{*}{MELD}     & \tnote{$\ddagger$}\; MM-DFN\cite{MM-DFN}  & 0.477 & 0.254 & 0.458 \\
                              & \tnote{$\dagger$}\; MMFA-RNN\cite{MMFA-RNN} & 0.500 & 0.243 & 0.446\\ \cline{2-5}
                              & SpeechFormer++ & \textbf{0.510} & \textbf{0.273} & \textbf{0.470}  \\ \hline
    \multirow{3}{*}{Pitt}     & Monica\cite{Monica} & 0.740 & 0.740 & 0.745  \\
                              & \tnote{$\dagger$}\; Makiuchi\cite{Makiuchi} & 0.794 & 0.791 & 0.774  \\ \cline{2-5}
                              & SpeechFormer++ & \textbf{0.813} & \textbf{0.816} & \textbf{0.808}  \\ \hline
    \multirow{4}{*}{DAIC-WOZ} & SIMSIAM-S\cite{daic_use_hubert} & 0.703 & - & -  \\
                              & \tnote{$\dagger$}\; FVTC-CNN\cite{FVTC-CNN} & 0.714 & 0.703 & 0.694  \\
                              & \tnote{$\ddagger$}\; EmoAudioNet\cite{EmoAudioNet} & 0.686 & 0.701 & 0.676  \\ \cline{2-5}
                              & SpeechFormer++ & \textbf{0.771} & \textbf{0.726} & \textbf{0.709}  \\ \hline
    \end{tabular}
    \begin{tablenotes}
        \footnotesize
        \item[$\dagger$] Source code is not provided. We reproduce the method by ourselves and replace the input features with HuBERT.
        \item[$\ddagger$] We reproduce the method using the code provided in the original paper and replace the input features with HuBERT.
    \end{tablenotes}
    \end{threeparttable}
\end{table}

\subsection{Ablation Study}
In this section, we conduct a comprehensive ablation study on the four corpora to determine the role of each module in our SpeechFormer++. All ablation studies are implemented in the same configuration, except for the module under investigation. In addition, we investigate the sensitivity of SpeechFormer++ to the statistical phone and word durations. Finally, we compare SpeechFormer++ with finetuning of HuBERT to verify the importance of the downstream model.

\subsubsection{Effectiveness of Unit Encoder} We first replace the unit encoder with the standard Transformer encoder, which means each layer in the modified model always applies full attention among all the acoustic tokens. The merging blocks are preserved in the modified model. Since the word encoder is used to enhance the unit encoder, we also remove the word encoder when the unit encoder is disabled. The results are reported in Table~\ref{tab_9}. The computational complexity is also listed for comparison. Supported by the unit encoder, SpeechFormer++ performs better than its counterpart by +2.77\% (+3.03\%) WA (UA) on IEMOCAP, +2.84\% WF1 on MELD, +2.14\% (+1.62\%) WA (UA) on pitt and +4.26\% MF1 on DAIC-WOZ. Generally, it demonstrates that modeling structure-based acoustic information under the instruction of the characteristics of speech can distinctly improve the performance of recognition. In addition, the attention mechanism in the proposed unit encoder scales linearly with the sequence length, which allows our SpeechFormer++ to achieve better performance at lower computational cost.

\begin{table}[!t]
    \caption{Ablation study on unit encoder (UE).\\Gain indicates the relative improvement (+) or reduction (-)}
    \label{tab_9}
    \centering
    \setlength{\tabcolsep}{3pt}{
    \begin{tabular}{c||cc||c||cc||c}
    \hline
    \multirow{2}{*}{UE}  & \multicolumn{2}{c||}{IEMOCAP}  & MELD  & \multicolumn{2}{c||}{Pitt}  & DAIC-WOZ   \\
         & WA & UA & WF1 & WA & UA & MF1 \\ \hline
    w/o  & 0.686  & 0.694  & 0.457 & 0.796  & 0.803  & 0.680  \\
    w/   & \textbf{0.705}  & \textbf{0.715}  & \textbf{0.470} & \textbf{0.813}  & \textbf{0.816}  & \textbf{0.709} \\ \hline
    Gain & +2.77\% & +3.03\% & +2.84\% & +2.14\% & +1.62\% & +4.26\% \\ \hline
    \multicolumn{7}{c}{FOLPs} \\ \hline
    w/o  & \multicolumn{2}{c||}{6.77G} & 4.55G & \multicolumn{2}{c||}{6.81G} & 9.09G \\ 
    w/  & \multicolumn{2}{c||}{6.55G} & 4.51G & \multicolumn{2}{c||}{6.58G} & 8.53G \\ \hline
    Gain & \multicolumn{2}{c||}{-3.25\%} & -0.88\% & \multicolumn{2}{c||}{-3.38\%} & -6.16\% \\ \hline
    \end{tabular}}
\end{table}

\begin{table}[!t]
    \caption{Ablation study on word encoder (WE).\\Gain indicates the relative improvement (+) or reduction (-)}
    \label{tab_10}
    \centering
    \setlength{\tabcolsep}{3pt}{
    \begin{tabular}{c||cc||c||cc||c}
    \hline
    \multirow{2}{*}{WE}  & \multicolumn{2}{c||}{IEMOCAP}  & MELD  & \multicolumn{2}{c||}{Pitt}  & DAIC-WOZ   \\
         & WA & UA & WF1 & WA & UA & MF1 \\ \hline
    w/o  & 0.701  & 0.709  & 0.464 & 0.806  & 0.805  & 0.679  \\
    w/   & \textbf{0.705}  & \textbf{0.715}  & \textbf{0.470} & \textbf{0.813}  & \textbf{0.816}  & \textbf{0.709} \\\hline
    Gain & +0.57\% & +0.85\% & +1.29\% & +0.87\% & +1.37\% & +4.42\% \\ \hline
    \multicolumn{7}{c}{FOLPs} \\ \hline
    w/o  & \multicolumn{2}{c||}{6.08G} & 4.18G & \multicolumn{2}{c||}{6.12G} & 7.93G \\ 
    w/  & \multicolumn{2}{c||}{6.55G} & 4.51G & \multicolumn{2}{c||}{6.58G} & 8.53G \\ \hline
    Gain & \multicolumn{2}{c||}{+7.73\%} & +7.89\% & \multicolumn{2}{c||}{+7.52\%} & +7.57\% \\ \hline
    \end{tabular}}
\end{table}

\subsubsection{Effectiveness of Word Encoder} To verify the potency of the word encoder, we discard all learnable word tokens in SpeechFormer++ such that the unit encoder can only perceive the local information within the window. The results are summarized in Table~\ref{tab_10}. The performance of the counterpart is weaker than SpeechFormer++ on four corpora, especially for IEMOCAP (0.696 vs. 0.705 in WA), MELD (0.464 vs. 0.470 in WF1), and DAIC-WOZ (0.672 vs. 0.709 in MF1). This indicates that the coarse-gained information cannot be neglected even if the fine-gained feature is effective. The word encoder presents an efficient way to help each unit encoder consider the coarse-gained information when modeling local segments. Note that the parameters in the word encoder and unit encoder are shared, enabling the introduction of the word encoder to not increase the model size but only the computational cost. In particular, the relative increments at FLOPs are +7.73\%, +7.89\%, +7.52\% and +7.57\% on IEMOCAP, MELD, Pitt and DAIC-WOZ, respectively.

\begin{table}[!t]
    \caption{Ablation study on merging Block (MB).\\Gain indicates the relative improvement (+) or reduction (-)}
    \label{tab_11}
    \centering
    \setlength{\tabcolsep}{3pt}{
    \begin{tabular}{c||cc||c||cc||c}
    \hline
    \multirow{2}{*}{MB}  & \multicolumn{2}{c||}{IEMOCAP}  & MELD  & \multicolumn{2}{c||}{Pitt}  & DAIC-WOZ   \\
         & WA & UA & WF1 & WA & UA & MF1 \\ \hline
    w/o  & 0.696  & 0.708  & 0.464 & 0.810  & 0.813  & 0.672  \\
    w/   & \textbf{0.705}  & \textbf{0.715}  & \textbf{0.470} & \textbf{0.813}  & \textbf{0.816}  & \textbf{0.709} \\\hline
    Gain & +1.29\% & +0.99\% & +1.29\% & +0.37\% & +0.37\% & +5.51\% \\ \hline
    \multicolumn{7}{c}{FOLPs} \\ \hline
    w/o  & \multicolumn{2}{c||}{22.17G} & 15.05G & \multicolumn{2}{c||}{22.31G} & 29.32G \\ 
    w/  & \multicolumn{2}{c||}{6.55G} & 4.51G & \multicolumn{2}{c||}{6.58G} & 8.53G \\ \hline
    Gain & \multicolumn{2}{c||}{-70.46\%} & -70.03\% & \multicolumn{2}{c||}{-70.51\%} & -70.91\% \\ \hline
    \end{tabular}}
\end{table}

\subsubsection{Effectiveness of Merging Block} To analyze the indispensability of the merging block, we implement a modified model with solely the unit encoder and the word encoder, where the number of input tokens is the same for each layer. In SpeechFormer++, each time the acoustic feature goes through a merging block, the number of tokens is greatly reduced, which significantly reduces the computational burden for the following layers. More concretely, as shown in Table~\ref{tab_11}, the computational complexity of SpeechFormer++ is reduced by 70.46\% on IEMOCAP, 70.03\% on MELD, 70.51\% on Pitt and 70.91\% on DAIC-WOZ compared to the counterpart. The model size is increased by merely 4.95\% over the Transformer. Furthermore, the presence of the merging block brings +0.37\%$\sim$+5.51\% relative gains over four corpora, which further confirms the necessity of the merging block.

\subsubsection{Sensitivity to the Statistical Durations}

\begin{figure}[t]
\centering
\includegraphics[width=3.4in]{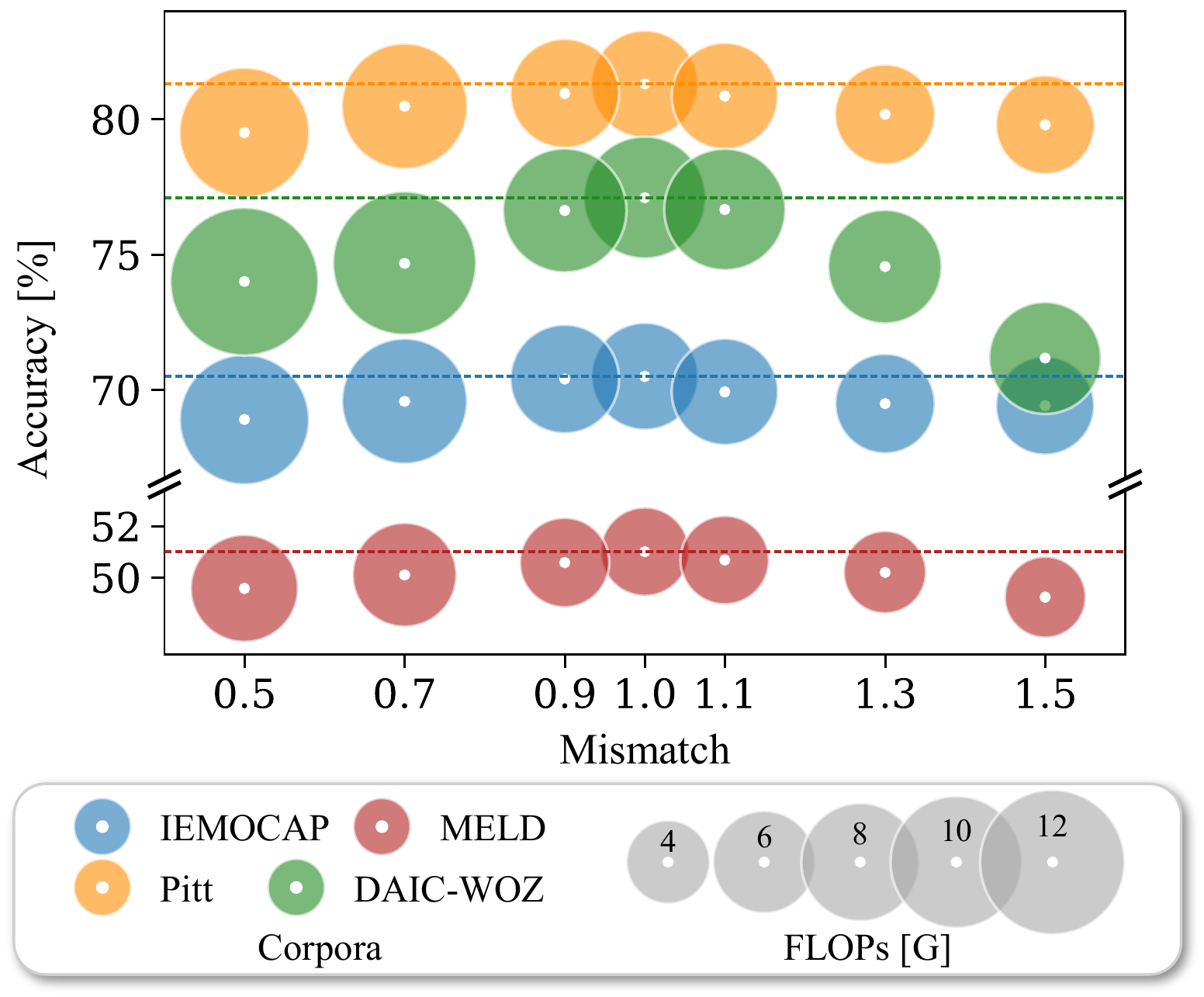}
\caption{Performance of SpeechFormer++ under different degrees of mismatch. The area of bubble is proportional to FLOPs. The color of bubble represents different corpora. The position of bubble's center represents the model accuracy. The dashed line indicates the accuracy of the baseline SpeechFormer++ using the statistics described in Section~\ref{statistics}.}
\label{fig_7}
\end{figure}

SpeechFormer++ is performed under the instruction of the statistical durations of speech. To investigate the sensitivity of SpeechFormer++ to the statistical phone and word durations, we intentionally set the phone and word durations longer or shorter than the statistics described in Section~\ref{statistics}. The experimental results are shown in Fig.~\ref{fig_7}. When the value of the x-axis $mismatch$ in Fig.~\ref{fig_7} is larger than 1, the durations of phone and word used in the system are $mismatch$ times longer than the respective statistics. On the contrary, if the $mismatch$ is less than 1, the durations used are shorter than the statistical durations. The durations are consistent with the statistics only if the $mismatch$ is equal to 1. The durations determine the window size in the encoder and the merging scale in the merging block, which further impact the performance and computational complexity. As shown in Fig.~\ref{fig_7}, the FLOPs gradually decreases when the durations used increase. The accuracy of SpeechFormer++ remains generally robust when the $mismatch$ lies between 0.9 and 1.1, suggesting that we can apply SpeechFormer++ directly to other English datasets with similar statistics. When the $mismatch$ is larger than 1.3 or less than 0.7, the performance starts to break, especially on DAIC-WOZ. These results suggest that we should recalculate the duration of each speech unit when processing different languages or language dialects.

\subsubsection{Comparison to Finetuning of Pretrained Model}
When the input feature is obtained from an already pretrained model (HuBERT in this paper), the proposed SpeechFormer++ can be viewed as a downstream model for the downstream task. To investigate the importance of the downstream model, we conduct experiments to compare the performances of finetuning the pretrained model with a simple MLP (3 dense layers) and learning further deep representation with SpeechFormer++. In addition, we implement Transformer and SpeechFormer++ with only 4 layers to investigate the impact of model size on small datasets. The experimental results in Table~\ref{tab_13} show that finetuning the pretrained model with the simple MLP obtains inferior results compared to the SpeechFormer++. The reason lies in that the pretrained model utilizes general self-supervised tasks, which do not consider the characteristics of speech. For results of SpeechFormer++ with 4 layers, we observe that it outperforms the 12 layers SpeechFormer++ on DAIC-WOZ. This is mainly because DAIC-WOZ is a small-scale dataset. Thus, only a small number of parameters are needed to fit the data. On the other three datasets, SpeechFormer++ with 12 layers delivers superior performance. Note that SpeechFormer++ also outperforms Transformer when both employ only 4 layers. Our work provides a new perspective on modeling speech signals. In this paper, we choose to use 12 layers of Transformer and SpeechFormer++ across four datasets merely for the sake of simple, straightforward and consistent comparisons. In real-world applications, the number of layers employed in each stage can be tuned for optimal performance according to the training dataset.

\begin{table}[!t]
    \caption{Performances of finetuning the pretrained model with a simple MLP and learning further deep representation with SpeechFormer++. MLP = Multilayer Perceptron
    }
    \label{tab_13}
    \centering
    \setlength{\tabcolsep}{3pt}{
    \begin{threeparttable}
    \begin{tabular}{c||cc||c||cc||c}
    \hline
    \multirow{2}{*}{Method}  & \multicolumn{2}{c||}{IEMOCAP} & MELD & \multicolumn{2}{c||}{Pitt} & DAIC-WOZ   \\
         & WA & UA & WF1 & WA & UA & MF1 \\ \hline
    \tnote{$\dagger$}\; MLP (FT)  & 0.677  & 0.689  & 0.455 & 0.793  & 0.789  &  0.676 \\
    \tnote{$\natural$}\; Transformer  & 0.680  & 0.699  & 0.457 & 0.798  & 0.793  & 0.665  \\ \hline
    \tnote{$\ddagger$}\; SpeechFormer++  & 0.695  & 0.702  & 0.468 & 0.805  & 0.805  & \textbf{0.746} \\
    SpeechFormer++ & \textbf{0.705}  & \textbf{0.715}  & \textbf{0.470} & \textbf{0.813}  & \textbf{0.816}  & 0.709 \\\hline
    \end{tabular}
    \begin{tablenotes}
        \footnotesize
        \item[$\dagger$] Finetune the pretrained HuBERT model with a simple MLP.
        \item[$\ddagger$] Each stage of SpeechFormer++ employs 1 layer, for a total of 4 layers.
        \item[$\natural$] Transformer with 4 layers.
    \end{tablenotes}
    \end{threeparttable}}
\end{table}

\subsection{Adopting Attention from Computer Vision}

\begin{table}[!t]
    \caption{Performances of adopting attention mechanisms from computer vision on four corpora. Window-based Swin and cluster-based BOAT algorithms are considered}
    \label{tab_12}
    \centering
    \setlength{\tabcolsep}{3pt}{
    \begin{tabular}{c||cc||c||cc||c}
    \hline
    \multirow{2}{*}{Method}  & \multicolumn{2}{c||}{IEMOCAP} & MELD & \multicolumn{2}{c||}{Pitt} & DAIC-WOZ   \\
         & WA & UA & WF1 & WA & UA & MF1 \\ \hline
    Swin-T\cite{Swin}  & 0.623  & 0.631  & 0.422 & 0.741  & 0.741  & 0.669  \\
    BOAT-T\cite{cluster_vit}  & 0.627  & 0.627  & 0.419 & 0.755  & 0.747  & 0.683 \\\hline
    Ours  & \textbf{0.705}  & \textbf{0.715}  & \textbf{0.470} & \textbf{0.813}  & \textbf{0.816}  & \textbf{0.709} \\\hline
    \multicolumn{7}{c}{FOLPs} \\ \hline
    Swin-T\cite{Swin} & \multicolumn{2}{c||}{11.50G} & 11.48G & \multicolumn{2}{c||}{11.50G} & 11.50G \\ 
    BOAT-T\cite{cluster_vit}  & \multicolumn{2}{c||}{15.01G} & 14.99G & \multicolumn{2}{c||}{15.01G} & 15.01G \\ \hline
    Ours  & \multicolumn{2}{c||}{6.55G} & 4.51G & \multicolumn{2}{c||}{6.58G} & 8.53G \\ \hline
    \end{tabular}}
\end{table}

We have confirmed that the standard full attention mechanism from NLP is unsuitable for the PSP task. Furthermore, we are interested in the performance of adopting attention methods from computer vision, which are optimized according to the characteristics of the images. Typically, the shifted window-based algorithm Swin-T\footnote{Codes of Swin: https://github.com/microsoft/Swin-Transformer} \cite{Swin} and the cluster-based algorithm BOAT-T\footnote{Codes of BOAT: https://github.com/mahaoyuHKU/pytorch-boat} \cite{cluster_vit} are considered in this paper. Note that we follow the same configurations as their original papers. The recognition results on four corpora are reported in Table~\ref{tab_12}. Computational costs are also provided for the sake of fair comparison. Unsurprisingly, the use of vision algorithms causes a significant drop in performance in the PSP tasks, as well as an increased computational burden. The results indicate that we cannot simply adapt the attention methods from other domains to speech, but instead, we need to make our own improvements based on the characteristics of the speech signal. Thus the proposed SpeechFormer++ presents a solution to fill the gap in the literature.

\subsection{Visualization of Attention Weights}

\begin{figure}[t]
\centering
\includegraphics[width=3.2in]{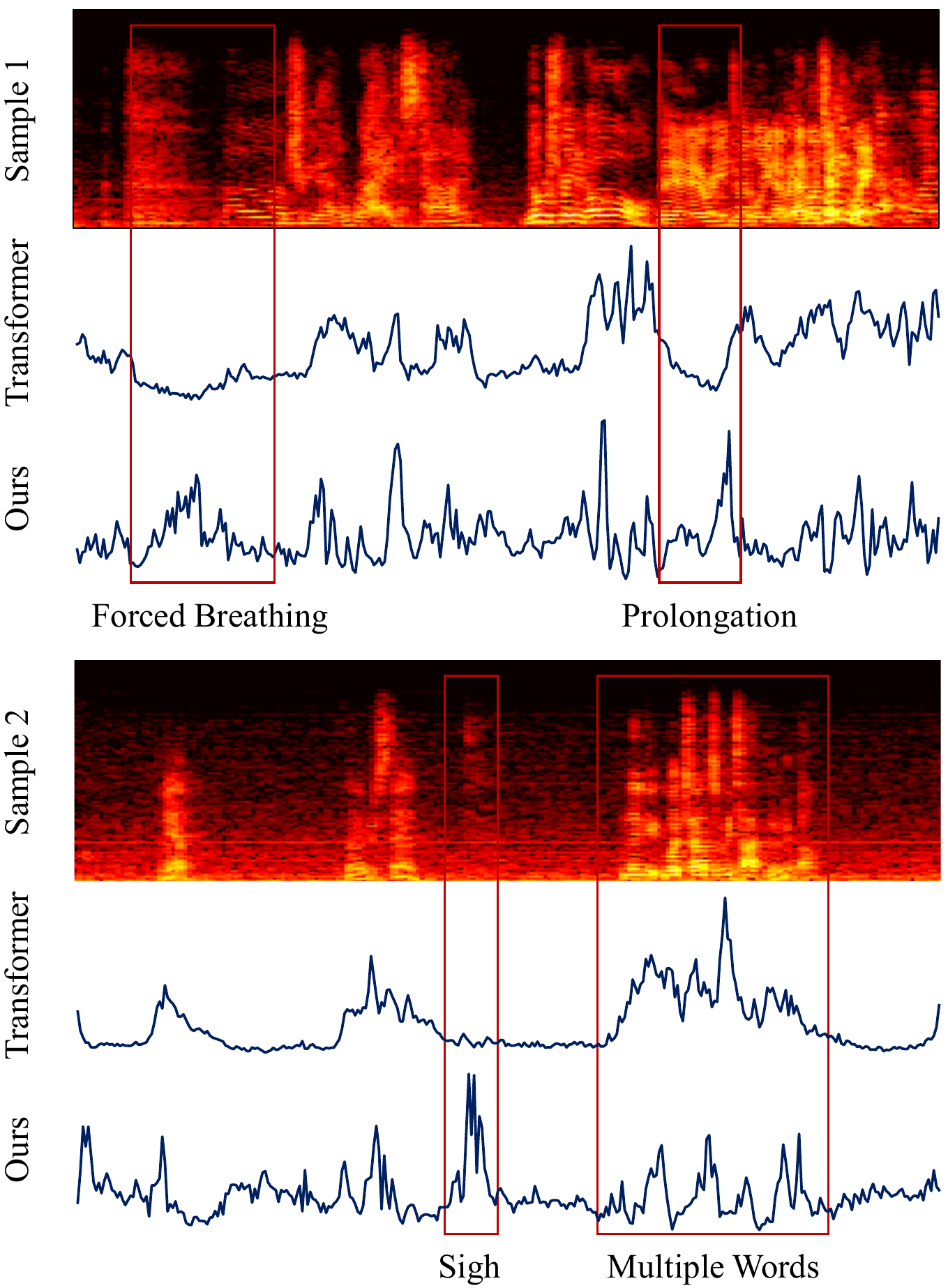}
\caption{Visualization of the normalized attention weights obtained by the Transformer and our proposed SpeechFormer++ on IEMOCAP. The spectrogram is given for better understanding and analysis. The notes are made by listening to the contents of the samples.}
\label{fig_6}
\end{figure}

From the experimental results discussed above, we conclude that SpeechFormer++ provides better results than the standard Transformer with less computational cost. It also outperforms the previous state-of-the-art approaches by a large margin on four commonly used corpora. To further understand the model and determine the reasons behind the improvements, we consider two utterance samples in IEMOCAP and compare the attention weights in Transformer and SpeechFormer++ by visualization. Here, the attention weights indicate the importance of each token in the model, which are obtained by adding up all the weights of the same value vector in MSA. Note that the merging block in SpeechFormer++ aggregates the acoustic tokens, resulting in a different number of input tokens in different layers and difficulties in comparison. For that reason, we visualize the attention weights in the first layer of SpeechFormer++ and Transformer, where the number of input tokens is the same for both models. The visualization results of the two utterance samples are illustrated in Fig.~\ref{fig_6}. The attention weights are limited to the range (0, 1) by softmax function. In addition, we manually mark out the content of the samples for better comprehension and analysis. For the first sample, forced breathing appears in the segment of the left bounding box, which is conducive for recognition. However, the attention weights in the Transformer indicate that the Transformer is not interested in that area and assigns relatively low weights to the corresponding tokens. Similarly, the right bounding box of the first sample is the prolongation of a particular word, which is essential for recognition but is omitted by Transformer. Our SpeechFormer++ is able to alleviate the above issues by assigning reasonable attention weights to the tokens in sample 1, where no informative content is neglected. For the second sample in Fig.~\ref{fig_6}, the left bounding box outlines an imperceptible sigh that is completely ignored by the Transformer and accurately captured by our SpeechFormer++. With this sigh signal, the model can accomplish the recognition effectively. The right bounding box of sample 2 shows a continuous utterance in which multiple words are spoken in quick succession. The attention weights in Transformer are relatively stable, while those in SpeechFormer++ fluctuate. This is because our method is capable of capturing more detailed information. In other words, SpeechFormer++ applies rapidly changing attention weights to model the fine-grained features in a cost-effective manner.

\section{Conclusion}
In this paper, we  exploit the potential of Transformer by considering the essence of the audio properties. We reveal the implicit relationships in speech and propose a structure-based framework, called SpeechFormer++, for paralinguistic speech processing. SpeechFormer++ takes the intra- and inter-unit features into account while preserving the coarse-grained information to further boost the performance. In addition, the merging blocks are applied to imitate the hierarchical structure in the speech signal. Experimental results on four speech-related corpora demonstrate that our method substantially surpasses the standard Transformer with respect to performance and efficiency. Additionally, the comparison to state-of-the-art also confirms the superiority of SpeechFormer++. In the future, we intend to make use of the lexical information and develop a unified textual-audio framework. In addition, we intend to consider the semantic information in SpeechFormer++ for solving speech recognition tasks.

\bibliographystyle{IEEEtran}
\bibliography{refs}

\begin{thebibliography}{10}
\providecommand{\url}[1]{#1}
\csname url@samestyle\endcsname
\providecommand{\newblock}{\relax}
\providecommand{\bibinfo}[2]{#2}
\providecommand{\BIBentrySTDinterwordspacing}{\spaceskip=0pt\relax}
\providecommand{\BIBentryALTinterwordstretchfactor}{4}
\providecommand{\BIBentryALTinterwordspacing}{\spaceskip=\fontdimen2\font plus
\BIBentryALTinterwordstretchfactor\fontdimen3\font minus
  \fontdimen4\font\relax}
\providecommand{\BIBforeignlanguage}[2]{{%
\expandafter\ifx\csname l@#1\endcsname\relax
\typeout{** WARNING: IEEEtran.bst: No hyphenation pattern has been}%
\typeout{** loaded for the language `#1'. Using the pattern for}%
\typeout{** the default language instead.}%
\else
\language=\csname l@#1\endcsname
\fi
#2}}
\providecommand{\BIBdecl}{\relax}
\BIBdecl

\bibitem{perception}
B.~Moore, L.~Tyler, and W.~Marslen-Wilson, ``Introduction. {The} perception of
  speech: from sound to meaning,'' \emph{Philosophical transactions of the
  Royal Society of London. Series B, Biological sciences}, vol. 363, no. 1493,
  pp. 917--921, Mar. 2008.

\bibitem{HMM1}
K.~Tokuda, T.~Kobayashi, and S.~Imai, ``Speech parameter generation from {HMM}
  using dynamic features,'' in \emph{IEEE International Conference on
  Acoustics, Speech, and Signal Processing}, vol.~1, 1995, pp. 660--663.

\bibitem{HMM2}
M.~Crouse, R.~Nowak, and R.~Baraniuk, ``Wavelet-based statistical signal
  processing using hidden {Markov} models,'' \emph{IEEE Transactions on Signal
  Processing}, vol.~46, no.~4, pp. 886--902, 1998.

\bibitem{HMM3}
B.~Schuller, G.~Rigoll, and M.~Lang, ``Hidden {Markov} model-based speech
  emotion recognition,'' in \emph{2003 International Conference on Multimedia
  and Expo. ICME '03. Proceedings}, vol.~1, 2003, pp. I--401.

\bibitem{Tree2}
J.~Cichosz and K.~Slot, ``Emotion recognition in speech signal using
  emotion-extracting binary decision trees,'' \emph{Proceedings of affective
  computing and intelligent interaction}, 2007.

\bibitem{Tree3}
L.~Yang, D.~Jiang, L.~He, E.~Pei, M.~C. Oveneke, and H.~Sahli, ``Decision tree
  based depression classification from audio video and language information,''
  in \emph{Proceedings of the 6th International Workshop on Audio/Visual
  Emotion Challenge}, ser. AVEC '16, 2016, pp. 89--96.

\bibitem{Boltzmann2}
A.~Stuhlsatz, C.~Meyer, F.~Eyben, T.~Zielke, G.~Meier, and B.~Schuller, ``Deep
  neural networks for acoustic emotion recognition: Raising the benchmarks,''
  in \emph{IEEE International Conference on Acoustics, Speech and Signal
  Processing}, 2011, pp. 5688--5691.

\bibitem{Haizhou3}
Z.~Wu, E.~S. Chng, and H.~Li, ``Conditional restricted {Boltzmann} machine for
  voice conversion,'' in \emph{2013 IEEE China Summit and International
  Conference on Signal and Information Processing}, 2013, pp. 104--108.

\bibitem{Boltzmann3}
K.~Poon-Feng, D.-Y. Huang, M.~Dong, and H.~Li, ``Acoustic emotion recognition
  based on fusion of multiple feature-dependent deep {Boltzmann} machines,'' in
  \emph{The 9th International Symposium on Chinese Spoken Language Processing},
  2014, pp. 584--588.

\bibitem{GCNN}
T.~Warnita, N.~Inoue, and K.~Shinoda, ``Detecting {Alzheimer}’s disease using
  gated convolutional neural network from audio data,'' in \emph{Interspeech},
  2018, pp. 1706--1710.

\bibitem{Makiuchi}
M.~R. Makiuchi, T.~Warnita, N.~Inoue, K.~Shinoda, M.~Yoshimura, M.~Kitazawa,
  K.~Funaki, Y.~Eguchi, and T.~Kishimoto, ``Speech paralinguistic approach for
  detecting dementia using gated convolutional neural network,'' \emph{IEICE
  Transactions on Information and Systems}, vol. 104, no.~11, pp. 1930--1940,
  2021.

\bibitem{EmoAudioNet}
A.~Othmani, D.~Kadoch, K.~Bentounes, E.~Rejaibi, R.~Alfred, and A.~Hadid,
  ``Towards robust deep neural networks for affect and depression recognition
  from speech,'' in \emph{Pattern Recognition. ICPR International Workshops and
  Challenges}, 2021, pp. 5--19.

\bibitem{FVTC-CNN}
Z.~Huang, J.~Epps, and D.~Joachim, ``Exploiting vocal tract coordination using
  dilated {CNNS} for depression detection in naturalistic environments,'' in
  \emph{IEEE International Conference on Acoustics, Speech and Signal
  Processing}, 2020, pp. 6549--6553.

\bibitem{stc}
L.~Guo, L.~Wang, C.~Xu, J.~Dang, E.~S. Chng, and H.~Li, ``Representation
  learning with spectro-temporal-channel attention for speech emotion
  recognition,'' in \emph{IEEE International Conference on Acoustics, Speech
  and Signal Processing}, 2021, pp. 6304--6308.

\bibitem{ISNet}
W.~Fan, X.~Xu, B.~Cai, and X.~Xing, ``{ISNet}: Individual standardization
  network for speech emotion recognition,'' \emph{IEEE/ACM Transactions on
  Audio, Speech, and Language Processing}, pp. 1--1, 2022.

\bibitem{fan2020}
W.~Fan, X.~Xu, X.~Xing, and D.~Huang, ``Adaptive domain-aware representation
  learning for speech emotion recognition,'' in \emph{Interspeech}, 2020, pp.
  4089--4093.

\bibitem{LSTM-GIN}
J.~Liu and H.~Wang, ``Graph isomorphism network for speech emotion
  recognition,'' in \emph{Interspeech}, 2021, pp. 3405--3409.

\bibitem{MM-DFN}
D.~Hu, X.~Hou, L.~Wei, L.~Jiang, and Y.~Mo, ``{MM-DFN}: Multimodal dynamic
  fusion network for emotion recognition in conversations,'' in \emph{IEEE
  International Conference on Acoustics, Speech and Signal Processing}, 2022,
  pp. 7037--7041.

\bibitem{MMFA-RNN}
N.-H. Ho, H.-J. Yang, S.-H. Kim, and G.~Lee, ``Multimodal approach of speech
  emotion recognition using multi-level multi-head fusion attention-based
  recurrent neural network,'' \emph{IEEE Access}, vol.~8, pp. 61\,672--61\,686,
  2020.

\bibitem{Autoencoder}
F.~Bertini, D.~Allevi, G.~Lutero, L.~Calzà, and D.~Montesi, ``An automatic
  {Alzheimer}’s disease classifier based on spontaneous spoken {English},''
  \emph{Computer Speech \& Language}, vol.~72, p. 101298, 2022.

\bibitem{SER_RNN1}
J.~Lee and I.~Tashev, ``High-level feature representation using recurrent
  neural network for speech emotion recognition,'' in \emph{Interspeech}, 2015,
  pp. 1537--1540.

\bibitem{Tao1}
L.~Chao, J.~Tao, M.~Yang, Y.~Li, and Z.~Wen, ``Long short term memory recurrent
  neural network based multimodal dimensional emotion recognition,'' in
  \emph{Proceedings of the 5th International Workshop on Audio/Visual Emotion
  Challenge}, ser. AVEC '15, 2015, pp. 65--72.

\bibitem{tao3}
J.~Huang, Y.~Li, J.~Tao, Z.~Lian, M.~Niu, and M.~Yang, ``Multimodal continuous
  emotion recognition with data augmentation using recurrent neural networks,''
  in \emph{Proceedings of the 2018 on Audio/Visual Emotion Challenge and
  Workshop}, ser. AVEC'18, 2018, pp. 57--64.

\bibitem{Romain1}
L.~Perotin, R.~Serizel, E.~Vincent, and A.~Guérin, ``Multichannel speech
  separation with recurrent neural networks from high-order ambisonics
  recordings,'' in \emph{IEEE International Conference on Acoustics, Speech and
  Signal Processing}, 2018, pp. 36--40.

\bibitem{SER_RNN3}
S.~T. Rajamani, K.~T. Rajamani, A.~Mallol-Ragolta, S.~Liu, and B.~Schuller, ``A
  novel attention-based gated recurrent unit and its efficacy in speech emotion
  recognition,'' in \emph{IEEE International Conference on Acoustics, Speech
  and Signal Processing}, 2021, pp. 6294--6298.

\bibitem{Transformer}
A.~Vaswani, N.~Shazeer, N.~Parmar, J.~Uszkoreit, L.~Jones, A.~N. Gomez,
  L.~Kaiser, and I.~Polosukhin, ``Attention is all you need,'' in
  \emph{Proceedings of the 31st International Conference on Neural Information
  Processing Systems}, 2017, pp. 5998--6008.

\bibitem{ViT}
A.~Dosovitskiy, L.~Beyer, A.~Kolesnikov, D.~Weissenborn, X.~Zhai,
  T.~Unterthiner, M.~Dehghani, M.~Minderer, G.~Heigold, S.~Gelly, J.~Uszkoreit,
  and N.~Houlsby, ``An image is worth 16x16 words: Transformers for image
  recognition at scale,'' in \emph{International Conference on Learning
  Representations}, 2021.

\bibitem{Swin}
Z.~Liu, Y.~Lin, Y.~Cao, H.~Hu, Y.~Wei, Z.~Zhang, S.~Lin, and B.~Guo, ``Swin
  transformer: Hierarchical vision transformer using shifted windows,''
  \emph{International Conference on Computer Vision}, 2021.

\bibitem{cluster_vit}
T.~Yu, G.~Zhao, P.~Li, and Y.~Yu, ``{BOAT}: Bilateral local attention vision
  transformer,'' \emph{arXiv preprint arXiv:2201.13027}, 2022.

\bibitem{cluster_CVPR}
W.~Zeng, S.~Jin, W.~Liu, C.~Qian, P.~Luo, W.~Ouyang, and X.~Wang, ``Not all
  tokens are equal: Human-centric visual analysis via token clustering
  transformer,'' in \emph{Proceedings of the IEEE/CVF Conference on Computer
  Vision and Pattern Recognition}, June 2022, pp. 11\,101--11\,111.

\bibitem{ViT2}
Q.~Zhang and Y.-B. Yang, ``{ResT}: An efficient transformer for visual
  recognition,'' in \emph{Advances in Neural Information Processing Systems},
  vol.~34, 2021, pp. 15\,475--15\,485.

\bibitem{drop_vit}
Y.~Rao, W.~Zhao, B.~Liu, J.~Lu, J.~Zhou, and C.-J. Hsieh, ``{DynamicViT}:
  Efficient vision transformers with dynamic token sparsification,'' in
  \emph{Advances in Neural Information Processing Systems}, vol.~34, 2021, pp.
  13\,937--13\,949.

\bibitem{Monica}
G.~M. Monica and M.~T. Rafael, ``A comparison of feature-based classifiers
  and transfer learning approaches for cognitive impairment recognition
  in language,'' in \emph{Artificial Intelligence in Neuroscience: Affective
  Analysis and Health Applications}, 2022, pp. 426--435.

\bibitem{ksT}
W.~Chen, X.~Xing, X.~Xu, J.~Yang, and J.~Pang, ``Key-sparse transformer for
  multimodal speech emotion recognition,'' in \emph{IEEE International
  Conference on Acoustics, Speech and Signal Processing}, 2022, pp. 6897--6901.

\bibitem{speech_use_trans2}
X.~Wang, M.~Wang, W.~Qi, W.~Su, X.~Wang, and H.~Zhou, ``A novel end-to-end
  speech emotion recognition network with stacked transformer layers,'' in
  \emph{IEEE International Conference on Acoustics, Speech and Signal
  Processing}, 2021, pp. 6289--6293.

\bibitem{speech_use_trans3}
K.~Chen, X.~Du, B.~Zhu, Z.~Ma, T.~Berg-Kirkpatrick, and S.~Dubnov, ``{HTS-AT}:
  A hierarchical token-semantic audio transformer for sound classification and
  detection,'' in \emph{IEEE International Conference on Acoustics, Speech and
  Signal Processing}, 2022, pp. 646--650.

\bibitem{ctnet}
Z.~Lian, B.~Liu, and J.~Tao, ``{CTNet}: Conversational transformer network for
  emotion recognition,'' \emph{IEEE/ACM Transactions on Audio, Speech, and
  Language Processing}, vol.~29, pp. 985--1000, 2021.

\bibitem{Tao2}
J.~Huang, J.~Tao, B.~Liu, Z.~Lian, and M.~Niu, ``Multimodal transformer fusion
  for continuous emotion recognition,'' in \emph{IEEE International Conference
  on Acoustics, Speech and Signal Processing}, 2020, pp. 3507--3511.

\bibitem{speechformer}
W.~Chen, X.~Xing, X.~Xu, J.~Pang, and L.~Du, ``{SpeechFormer}: A hierarchical
  efficient framework incorporating the characteristics of speech,'' in
  \emph{Interspeech}, 2022, pp. 346--350.

\bibitem{TASLP_QA}
Y.~Cui, T.~Liu, W.~Che, Z.~Chen, and S.~Wang, ``Teaching machines to read,
  answer and explain,'' \emph{IEEE/ACM Transactions on Audio, Speech, and
  Language Processing}, vol.~30, pp. 1483--1492, 2022.

\bibitem{nlp_use_trans_ner3}
J.~Yu, J.~Jiang, L.~Yang, and R.~Xia, ``Improving multimodal named entity
  recognition via entity span detection with unified multimodal transformer,''
  in \emph{Proceedings of the 58th Annual Meeting of the Association for
  Computational Linguistics}.\hskip 1em plus 0.5em minus 0.4em\relax
  Association for Computational Linguistics, 2020.

\bibitem{nlp_use_trans_nli1}
M.~Guo, Y.~Zhang, and T.~Liu, ``Gaussian transformer: A lightweight approach
  for natural language inference,'' \emph{Proceedings of the AAAI Conference on
  Artificial Intelligence}, vol.~33, no.~01, pp. 6489--6496, Jul. 2019.

\bibitem{TASLP_sentence_embedding}
B.~Wang and C.-C.~J. Kuo, ``{SBERT-WK}: A sentence embedding method by
  dissecting {BERT}-based word models,'' \emph{IEEE/ACM Transactions on Audio,
  Speech, and Language Processing}, vol.~28, pp. 2146--2157, 2020.

\bibitem{nlp_use_trans_dc3}
C.~Wu, F.~Wu, T.~Qi, and Y.~Huang, ``{Hi-Transformer}: Hierarchical interactive
  transformer for efficient and effective long document modeling,'' in
  \emph{Proceedings of the 59th Annual Meeting of the Association for
  Computational Linguistics and the 11th International Joint Conference on
  Natural Language Processing}, Aug. 2021, pp. 848--853.

\bibitem{deepvit}
D.~Zhou, B.~Kang, X.~Jin, L.~Yang, X.~Lian, Z.~Jiang, Q.~Hou, and J.~Feng,
  ``{DeepViT}: Towards deeper vision transformer,'' \emph{arXiv preprint
  arXiv:2103.11886}, 2021.

\bibitem{speech_use_trans1}
A.~Nediyanchath, P.~Paramasivam, and P.~Yenigalla, ``Multi-head attention for
  speech emotion recognition with auxiliary learning of gender recognition,''
  in \emph{IEEE International Conference on Acoustics, Speech and Signal
  Processing}, 2020, pp. 7179--7183.

\bibitem{Saliency}
Y.-T. Wu, J.-L. Li, and C.-C. Lee, ``An audio-saliency masking transformer for
  audio emotion classification in movies,'' in \emph{IEEE International
  Conference on Acoustics, Speech and Signal Processing}, 2022, pp. 4813--4817.

\bibitem{ad_trans1}
L.~Ilias, D.~Askounis, and J.~Psarras, ``Detecting dementia from speech and
  transcripts using transformers,'' \emph{arXiv preprint arXiv:2110.14769},
  2021.

\bibitem{ad_trans2}
Y.~Zhu, A.~Obyat, X.~Liang, J.~A. Batsis, and R.~M. Roth, ``{WavBERT}:
  Exploiting semantic and non-semantic speech using wav2vec and {BERT} for
  dementia detection,'' in \emph{Interspeech}, 2021, pp. 3790--3794.

\bibitem{depression_trans1}
H.~Sun, J.~Liu, S.~Chai, Z.~Qiu, L.~Lin, X.~Huang, and Y.~Chen, ``Multi-modal
  adaptive fusion transformer network for the estimation of depression level,''
  \emph{Sensors}, vol.~21, no.~14, p. 4764, Jul 2021.

\bibitem{wav2vec}
S.~Schneider, A.~Baevski, R.~Collobert, and M.~Auli, ``wav2vec: Unsupervised
  pre-training for speech recognition,'' in \emph{Interspeech}, 2019, pp.
  3465--3469.

\bibitem{wav2vec2}
A.~Baevski, Y.~Zhou, A.~Mohamed, and M.~Auli, ``wav2vec 2.0: A framework for
  self-supervised learning of speech representations,'' in \emph{Advances in
  Neural Information Processing Systems}, vol.~33, 2020, pp. 12\,449--12\,460.

\bibitem{hubert}
W.-N. Hsu, B.~Bolte, Y.-H.~H. Tsai, K.~Lakhotia, R.~Salakhutdinov, and
  A.~Mohamed, ``{HuBERT}: Self-supervised speech representation learning by
  masked prediction of hidden units,'' \emph{IEEE/ACM Transactions on Audio,
  Speech, and Language Processing}, vol.~29, pp. 3451--3460, 2021.

\bibitem{CA-MSER}
H.~Zou, Y.~Si, C.~Chen, D.~Rajan, and E.~S. Chng, ``Speech emotion recognition
  with co-attention based multi-level acoustic information,'' in \emph{IEEE
  International Conference on Acoustics, Speech and Signal Processing}, 2022,
  pp. 7367--7371.

\bibitem{reviewer3_1}
J.~Wagner, A.~Triantafyllopoulos, H.~Wierstorf, M.~Schmitt, F.~Burkhardt,
  F.~Eyben, and B.~W. Schuller, ``Dawn of the transformer era in speech emotion
  recognition: {Closing} the valence gap,'' \emph{arXiv preprint
  arXiv:2203.07378}, 2022.

\bibitem{sota_w2v2_daic_1}
Y.~Guo, C.~Zhu, S.~Hao, and R.~Hong, ``A topic-attentive transformer-based
  model for multimodal depression detection,'' \emph{arXiv preprint
  arXiv:2206.13256}, 2022.

\bibitem{sota_w2v2_meld_1}
M.~Sharma, ``Multi-lingual multi-task speech emotion recognition using wav2vec
  2.0,'' in \emph{IEEE International Conference on Acoustics, Speech and Signal
  Processing}, 2022, pp. 6907--6911.

\bibitem{TaoJianhua_asr}
Z.~Tian, J.~Yi, J.~Tao, Y.~Bai, and Z.~Wen, ``Self-attention transducers for
  end-to-end speech recognition,'' in \emph{Interspeech}, 2019, pp. 4395--4399.

\bibitem{reviewer2_2}
Y.~Wang, A.~Mohamed, D.~Le, C.~Liu, A.~Xiao, J.~Mahadeokar, H.~Huang,
  A.~Tjandra, X.~Zhang, F.~Zhang, C.~Fuegen, G.~Zweig, and M.~L. Seltzer,
  ``Transformer-based acoustic modeling for hybrid speech recognition,'' in
  \emph{IEEE International Conference on Acoustics, Speech and Signal
  Processing}, 2020, pp. 6874--6878.

\bibitem{reviewer2_1}
A.~Gulati, J.~Qin, C.-C. Chiu, N.~Parmar, Y.~Zhang, J.~Yu, W.~Han, S.~Wang,
  Z.~Zhang, Y.~Wu, and R.~Pang, ``{Conformer}: Convolution-augmented
  transformer for speech recognition,'' in \emph{Interspeech}, 2020, pp.
  5036--5040.

\bibitem{reviewer2_3}
Z.~Zhao, Z.~Bao, Z.~Zhang, J.~Deng, N.~Cummins, H.~Wang, J.~Tao, and
  B.~Schuller, ``Automatic assessment of depression from speech via a
  hierarchical attention transfer network and attention autoencoders,''
  \emph{IEEE Journal of Selected Topics in Signal Processing}, vol.~14, no.~2,
  pp. 423--434, 2020.

\bibitem{reviewer2_4}
Y.~Gu, K.~Yang, S.~Fu, S.~Chen, X.~Li, and I.~Marsic, ``Multimodal affective
  analysis using hierarchical attention strategy with word-level alignment,''
  in \emph{Proceedings of the conference. Association for Computational
  Linguistics. Meeting}, vol. 2018, 2018, p. 2225.

\bibitem{speech_nature1}
G.~Shen, R.~Lai, R.~Chen, Y.~Zhang, K.~Zhang, Q.~Han, and H.~Song, ``{WISE}:
  Word-level interaction-based multimodal fusion for speech emotion
  recognition.'' in \emph{Interspeech}, 2020, pp. 369--373.

\bibitem{speech_nature2}
H.~Li, W.~Ding, Z.~Wu, and Z.~Liu, ``Learning fine-grained cross modality
  excitement for speech emotion recognition,'' in \emph{Interspeech}, 2021, pp.
  3375--3379.

\bibitem{speech_nature3}
Y.~Wang, G.~Shen, Y.~Xu, J.~Li, and Z.~Zhao, ``Learning mutual correlation in
  multimodal transformer for speech emotion recognition,'' in
  \emph{Interspeech}, 2021, pp. 4518--4522.

\bibitem{gru}
K.~Cho, B.~van Merrienboer, C.~Gulcehre, F.~Bougares, H.~Schwenk, and
  Y.~Bengio, ``Learning phrase representations using {RNN} encoder-decoder for
  statistical machine translation,'' in \emph{Conference on Empirical Methods
  in Natural Language Processing}, 2014, pp. 1724--1734.

\bibitem{yoked}
M.~Keshishian, S.~Norman-Haignere, and N.~Mesgarani, ``Understanding adaptive,
  multiscale temporal integration in deep speech recognition systems,'' in
  \emph{Advances in Neural Information Processing Systems}, vol.~34, 2021, pp.
  24\,455--24\,467.

\bibitem{P2FA}
J.~Yuan and M.~Y. Liberman, ``Speaker identification on the scotus corpus,''
  \emph{Journal of the Acoustical Society of America}, vol. 123, pp.
  3878--3878, 2008.

\bibitem{LN}
J.~{Lei Ba}, J.~R. {Kiros}, and G.~E. {Hinton}, ``Layer normalization,''
  \emph{arXiv preprint arXiv:1607.06450}, 2016.

\bibitem{IEMOCAP}
C.~Busso, M.~Bulut, C.-C. Lee, A.~Kazemzadeh, E.~Mower, S.~Kim, J.~N. Chang,
  S.~Lee, and S.~S. Narayanan, ``{IEMOCAP: Interactive emotional dyadic motion
  capture database},'' \emph{Language Resources and Evaluation}, vol.~42,
  no.~4, pp. 335--359, 2008.

\bibitem{meld}
S.~Poria, D.~Hazarika, N.~Majumder, G.~Naik, E.~Cambria, and R.~Mihalcea,
  ``{MELD: A} multimodal multi-party dataset for emotion recognition in
  conversations,'' \emph{arXiv preprint arXiv:1810.02508}, 2019.

\bibitem{pitt}
J.~T. Becker, F.~Boiler, O.~L. Lopez, J.~Saxton, and K.~L. McGonigle, ``The
  natural history of {Alzheimer}'s disease: Description of study cohort and
  accuracy of diagnosis,'' \emph{Archives of Neurology}, vol.~51, no.~6, pp.
  585--594, 1994.

\bibitem{cookie}
K.~E. Goodglass~H, ``The {Boston} diagnostic aphasia examination,'' \emph{Lea
  \& Febinger, Philadelphia}, 1983.

\bibitem{avec}
J.~Gratch, R.~Artstein, G.~Lucas, G.~Stratou, S.~Scherer, A.~Nazarian, R.~Wood,
  J.~Boberg, D.~DeVault, S.~Marsella, D.~Traum, S.~Rizzo, and L.-P. Morency,
  ``The distress analysis interview corpus of human and computer interviews,''
  in \emph{Proceedings of the Ninth International Conference on Language
  Resources and Evaluation}, May 2014, pp. 3123--3128.

\bibitem{avec2017}
F.~Ringeval, B.~Schuller, M.~Valstar, J.~Gratch, R.~Cowie, S.~Scherer,
  S.~Mozgai, N.~Cummins, M.~Schmitt, and M.~Pantic, ``{AVEC} 2017: Real-life
  depression, and affect recognition workshop and challenge,'' in
  \emph{Proceedings of the 7th Annual Workshop on Audio/Visual Emotion
  Challenge}, ser. AVEC '17, 2017, pp. 3--9.

\bibitem{12layer}
J.~Shah, Y.~K. Singla, C.~Chen, and R.~R. Shah, ``What all do audio transformer
  models hear? {Probing} acoustic representations for language delivery and its
  structure,'' \emph{arXiv preprint arXiv:2101.00387}, 2021.

\bibitem{superb}
S.~wen Yang, P.-H. Chi, Y.-S. Chuang, C.-I.~J. Lai, K.~Lakhotia, Y.~Y. Lin,
  A.~T. Liu, J.~Shi, X.~Chang, G.-T. Lin, T.-H. Huang, W.-C. Tseng, K.~tik Lee,
  D.-R. Liu, Z.~Huang, S.~Dong, S.-W. Li, S.~Watanabe, A.~Mohamed, and
  H.~yi~Lee, ``{SUPERB: S}peech processing universal performance benchmark,''
  in \emph{Interspeech}, 2021, pp. 1194--1198.

\bibitem{ConGCN}
D.~Zhang, L.~Wu, C.~Sun, S.~Li, Q.~Zhu, and G.~Zhou, ``Modeling both context-
  and speaker-sensitive dependence for emotion detection in multi-speaker
  conversations.'' in \emph{International Joint Conference on Artificial
  Intelligence}, 2019, pp. 5415--5421.

\bibitem{pitt_use_w2v2}
P.~A. Pérez-Toro, P.~Klumpp, A.~Hernandez, T.~Arias, P.~Lillo, A.~Slachevsky,
  A.~M. García, M.~Schuster, A.~K. Maier, E.~Noeth, and J.~R. Orozco-Arroyave,
  ``{A}lzheimer's detection from {E}nglish to {S}panish using acoustic and
  linguistic embeddings,'' in \emph{Interspeech}, 2022, pp. 2483--2487.

\bibitem{Saidi}
A.~Saidi, S.~B. Othman, and S.~B. Saoud, ``Hybrid {CNN}-{SVM} classifier for
  efficient depression detection system,'' in \emph{International Conference on
  Advanced Systems and Emergent Technologies}, 2020, pp. 229--234.

\bibitem{Solieman}
H.~Solieman and E.~A. Pustozerov, ``The detection of depression using
  multimodal models based on text and voice quality features,'' in \emph{2021
  IEEE Conference of Russian Young Researchers in Electrical and Electronic
  Engineering}, 2021, pp. 1843--1848.

\bibitem{daic_use_hubert}
S.~H. Dumpala, C.~S. Sastry, R.~Uher, and S.~Oore, ``On combining global and
  localized self-supervised models of speech,'' in \emph{Interspeech}, 2022,
  pp. 3593--3597.

\end{thebibliography}


\vspace{11pt}
\vspace{-33pt}
\begin{IEEEbiography}[{\includegraphics[width=1in,height=1.25in,clip,keepaspectratio]{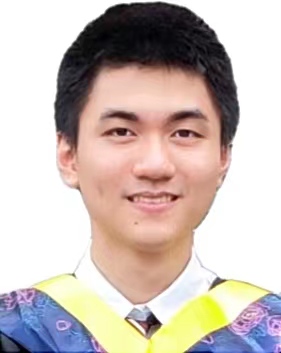}}]{Weidong Chen} received the B.E. degree in electronic science and technology from South China University of Technology, Guangzhou, China, in 2021. He is currently working toward the M.E. degree with the School of Electronic and Information Engineering, South China University of Technology. His research interests include speech emotion recognition, multimodal emotion recognition, and deep learning in speech processing.
\end{IEEEbiography}

\vspace{11pt}
\vspace{-33pt}
\begin{IEEEbiography}[{\includegraphics[width=1in,height=1.25in,clip,keepaspectratio]{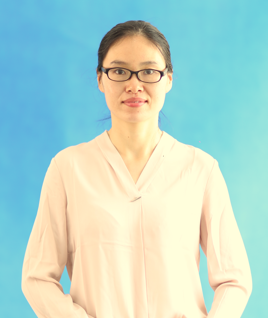}}]{Xiaofen Xing} (Member, IEEE) received the B.S., M.S., and Ph.D. degrees from South China University of Technology, Guangzhou, China, in 2001, 2004 and 2013, respectively. Since 2017, she has been an Associate Professor with the School of Electronic and Information Engineering, South China University of Technology. Her main research interests include speech emotion analysis, image/video processing, and human computer interaction.
\end{IEEEbiography}

\vspace{11pt}
\vspace{-33pt}
\begin{IEEEbiography}[{\includegraphics[width=1in,height=1.25in,clip,keepaspectratio]{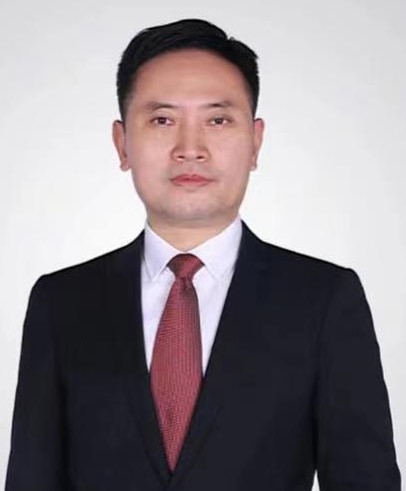}}]{Xiangmin Xu} (Senior Member, IEEE) received the Ph.D. degree from South China University of Technology, Guangzhou, China. He is currently a Full Professor with the School of Electronic and Information Engineering, South China University of Technology. His research interests include image/video processing, human computer interaction, computer vision, and machine learning.
\end{IEEEbiography}

\vspace{11pt}
\vspace{-33pt}
\begin{IEEEbiography}[{\includegraphics[width=1in,height=1.25in,clip,keepaspectratio]{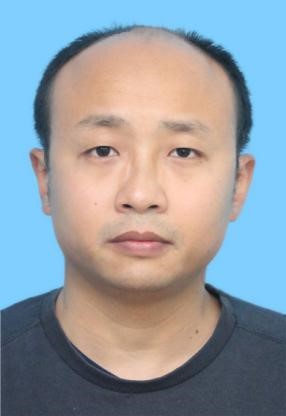}}]{Jianxin Pang} (Member, IEEE) received the B.E. and Ph.D. degrees from the University of Science and Technology of China, Hefei, China, in 2002 and 2008, respectively. He is a Senior Engineer and VP with UBTECH Robotics, Shenzhen, China. His research interests include image and video understanding, human–computer interaction, and intelligent robot.

\end{IEEEbiography}

\vspace{11pt}
\vspace{-33pt}
\begin{IEEEbiography}[{\includegraphics[width=1in,height=1.25in,clip,keepaspectratio]{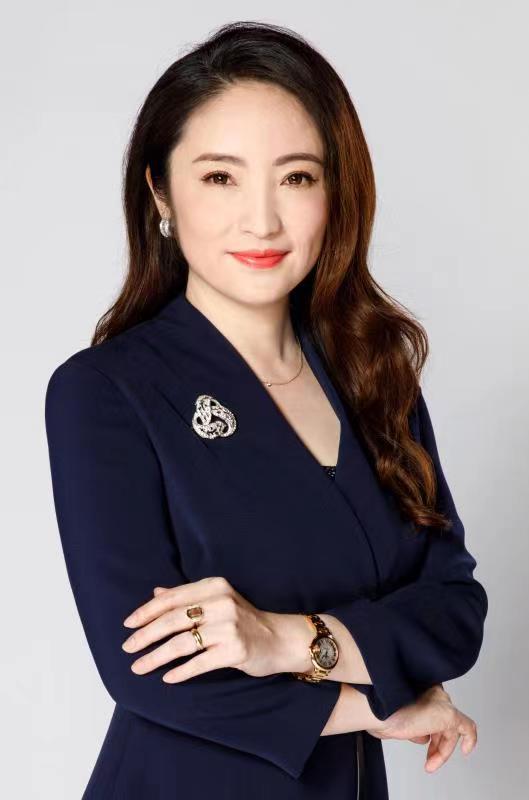}}]{Lan Du} received the Ph.D. degree from South China University of Technology, and worked as a Post-doctor at South China University of Technology. She has been hired as the adjunct professor of Renmin University of China, Communication University of China, Sun Yat-sen University, South China University of Technology and Jinan University. Her research interests include artificial intelligence.
\end{IEEEbiography}

\vfill

\end{document}